\documentclass[12pt]{article}
\usepackage{amsmath,amssymb,exscale,cancel}
\usepackage{slashed}

\usepackage{amsmath}
\usepackage{graphicx}
\usepackage{epsfig}
\usepackage{multicol}
\usepackage{color}
\usepackage{mathrsfs}
\usepackage{blindtext}
 \usepackage{fancyhdr}
\usepackage{hyperref}
\usepackage{cite}
\usepackage{mathtools}

\usepackage{floatrow}

\usepackage{graphicx}
\usepackage{sidecap}

\usepackage[latin1]{inputenc} 

\usepackage{multicol}  
 
 \usepackage{titlesec}
 
 \usepackage{rotating,slashed,xcolor,amsfonts,expdlist,charter}

\numberwithin{equation}{section}

\usepackage{xcolor}
\usepackage{sectsty}

\usepackage{mdframed}
\usepackage{titletoc}

\numberwithin{equation}{section}

\usepackage{xcolor}
\usepackage{sectsty}

\usepackage{mdframed}
\usepackage{titletoc}

\definecolor{secnum}{RGB}{13,151,225}
\definecolor{ptcbackground}{RGB}{212,237,252}
\definecolor{ptctitle}{RGB}{0,177,235}

\titlecontents{lsection}
  [5.8em]{\sffamily}
  {\color{secnum}\contentslabel{2.3em}\normalcolor}{}
  {\titlerule*[1000pc]{.}\contentspage\\\hspace*{-5.8em}\vspace*{5pt}%
    \color{white}\rule{\dimexpr\textwidth-15.5pt\relax}{1pt}}

\usepackage{hyperref}
\hypersetup{colorlinks,bookmarksopen,bookmarksnumbered,citecolor=rossos,
linkcolor=redy,pdfstartview=FitH,urlcolor=green-go}
\usepackage{slashed}

\definecolor{blus}{cmyk}{1,0.9,0,0.1}
\definecolor{verdes}{cmyk}{0.99,0,0.59,0.65}
\definecolor{rossos}{cmyk}{0,1,1,0.55}
\definecolor{redy}{cmyk}{0,1,1,0.7}
\definecolor{greeny}{cmyk}{0.99,0,0.59,0.98}
\definecolor{green-go}{cmyk}{0.79,0,0.59,0.5}

\usepackage{titlesec}

\def\tt{{\tiny \times}}

\newcommand{\beq}{\begin{equation}}
\newcommand{\eeq}{\end{equation}}

\def\hhref#1{\href{http://arxiv.org/abs/#1}{arXiv:#1}} % in bibliography

 \def\Lag{\mathscr{L}}
 
\newcommand{\tmtextbf}[1]{{\bfseries{#1}}}
\newcommand{\tmtextrm}[1]{{\rmfamily{#1}}}
\newcommand{\bp}{\bar M_P}

\def\be{\begin{equation}}
\def\ee{\end{equation}}
\def\ba{\begin{array} }

\def\bac{\begin{array} {c}}
\def\bacc{\begin{array} {cc}}
\def\baccc{\begin{array} {ccc}}
\def\bacccc{\begin{array} {cccc}}
\def\ea{\end{array}}
\def\bea{\begin{eqnarray}}
\def\eea{\end{eqnarray}}

\definecolor{red}{rgb}{1,0,0}

\def\psl{\hbox{\hbox{${p}$}}\kern-1.9mm{\hbox{${/}$}}}
\def\dsl{\hbox{\hbox{${\partial}$}}\kern-2.2mm{\hbox{${/}$}}}
\def\Dsl{\hbox{\hbox{${D}$}}\kern-2.6mm{\hbox{${/}$}}}

\newcommand{\gappeq}{{\rlap{{\raise}.5ex\text{\ensuremath{>}}}{{\lower}.5ex\text{\ensuremath{\sim}}}}}
\newcommand{\lappeq}{{\rlap{{\raise}.5ex\text{\ensuremath{<}}}{{\lower}.5ex\text{\ensuremath{\sim}}}}}
\newcommand{\I}{\tmtextrm{1{\kern}-.24em l}}

\begin{document}
\topmargin -1.0cm
\oddsidemargin 0.9cm
\evensidemargin -0.5cm

{\vspace{-1cm}}
\begin{center}

\vspace{-1cm}

 {\fontsize{19.8pt}{18pt}\selectfont \tmtextbf{ 
\color{rossos} \hspace{-0.9cm}    \\
On the Role of Gravity  in  Supercooled Phase Transitions 
 \hspace{-1.6cm}}} {\vspace{.5cm}}\\

%TO DO
%introduction   x
%conclusions x
%abstract x
%ordinamento refs x
%controllare coerenza simboli  x
%acronimi x
%logic and formulae x
%add refs using 
%  - JCAPedt x
%  - iNSPIREme x
%  - the relevant literature folders of "TFT" x
%check  English x
%print and read it again 

%remove blue and red
%submit to arXiv
%submit to journal

\vspace{1.3cm}

{\large{\bf   Filippo Cutrona {\rm and} Alberto Salvio}}

{\em  

\vspace{0.6cm}

Physics Department, University of Rome and INFN Tor Vergata, Italy\\

%
%\vspace{.4cm}
%
%%\today 
%
%\vspace{0.4cm}
% 
%
%
%\vspace{0.2cm}
%
 \vspace{1.5cm}

%\vspace{.3cm}
%
%
% 
%
%
}
%\vspace{0.cm}
%
%
\end{center}

%
% \begin{\large abstract}
% 
\noindent ---------------------------------------------------------------------------------------------------------------------------------
\begin{center}
{\bf \large Abstract}
\end{center}
\noindent  Theories with radiative mass generation and radiative symmetry breaking (RSB)  lead to first-order phase transitions (FOPTs) and the production of gravitational waves (GWs) as well as primordial black holes if supercooling lasts long enough. Here we investigate the role of gravity in supercooled FOPTs that occur in the RSB scenario. We use a model-independent approach that is valid when a quantity (which we explicitly determine and  is computable once the model is specified) is small enough. Eventually, the curvature of spacetime becomes relevant if the FOPT does not complete in the case in which gravity is ignored. We also include the effect on the FOPT of the non-minimal couplings  with coefficients of zero mass dimension.  Depending on such non-minimal couplings, gravity can enhance or reduce the region of parameter space where the phase transition completes. Observable consequences of the role of gravity on the FOPT that can be probed by current and future GW detectors are then worked out and we explore the possibility that the  background of GWs detected by pulsar timing arrays is produced by FOPTs in the RSB scenario. We also estimate the number of e-folds that elapse when the field playing the role of the order parameter is trapped in the false vacuum, which provides important information for cosmology.

\vspace{0.7cm}

\noindent---------------------------------------------------------------------------------------------------------------------------------

  \vspace{-0.9cm}
  
  \newpage 
\tableofcontents

\noindent

\vspace{0.5cm}

\noindent ---------------------------------------------------------------------------------------------------------------------------------

\section{Introduction}\label{Introduction}

GW astronomy has become an extremely active and exciting field of physics after the discovery of the GWs from  binary black hole and neutron star mergers~\cite{Abbott:2016blz,TheLIGOScientific:2016wyq1,LIGOScientific:2017ync}. A few years ago the interest in this field further increased after the detection of a background of GWs by pulsar timing arrays (PTAs): these include the North American Nanohertz Observatory for Gravitational Waves (NANOGrav), the Chinese Pulsar Timing Array (CPTA), the European Pulsar Timing Array (EPTA) and the Parkes Pulsar Timing Array (PPTA)~\cite{NANOGrav:2023gor,NANOGrav:2023hvm,Antoniadis:2023ott,Reardon:2023gzh,Xu:2023wog}. Moreover, the discovery of the GWs from  binary black hole mergers has revived the interest in primordial black holes as (part of) the observed dark matter abundance.
 
Quite generically,  supercooled first-order phase transitions (PTs) can lead to observable GWs and primordial black holes (see e.g.~\cite{Hawking:1982ga,Crawford:1982yz,Kodama:1982sf,Moss:1994iq,Khlopov:1999ys,Maggiore:2018sht,Kawana:2021tde,Liu:2021svg,Baker:2021sno,Kawana:2022olo,Athron:2023xlk,Lewicki:2023ioy,Gouttenoire:2023naa,Ahriche:2023jdq,Hashino:2025fse,Kierkla:2025vwp}). In these PTs the effective potential of the field $\chi$, whose vacuum-expectation value plays the role of the order parameter, below a certain value $T_c$ of the temperature $T$, develops a true vacuum (global minimum)  and  a false vacuum (local minimum) and $\chi$ remains trapped in the false vacuum for a long time. During this epoch  $T$ decreases exponentially.

 Generically, the more the FOPT is delayed, the more GWs and primordial black holes are formed. On flat spacetime, such PTs always  occur~\cite{Witten:1980ez,Salvio:2023qgb,Salvio:2023ext,Banerjee:2024cwv,Salvio:2024upo,Salvio:2026bco} when symmetries are broken and masses are generated mostly through radiative effects~\cite{Coleman:1973jx,Gildener:1976ih}. 
This RSB scenario also offers the possibility of explaining large mass hierarchies  as these are generated through dimensional transmutation~\cite{Gildener:1976ih} and is a quite predictive framework because only dimensionless parameters are allowed in the classical action. Also, in the RSB scenario and in the absence of gravity the PTs always feature a long period of supercooling~\cite{Witten:1980ez,Salvio:2023qgb}, when the field $\chi$ responsible for the PT is trapped in the false vacuum before the PT actually takes place. 
It is important to point out that RSB models may be either Standard Model (SM) extensions or  ``dark" sectors weakly coupled to the SM.

Refs.~\cite{Salvio:2023qgb,Salvio:2023ext,Banerjee:2024cwv} showed, working on flat spacetime, that if supercooling is strong enough a model-independent description of the PTs and the consequent production of gravitational waves and primordial black holes becomes possible in the perturbative RSB scenario. Supercooling generically leads to strong PTs~\cite{Salvio:2023qgb}.

The purpose of this paper is to extend these works and explore the role of gravity in supercooled PTs, considering the RSB scenario in a model-independent way. Understanding gravitational effects is needed in the region of the parameter space where, ignoring gravity,  the PT does not complete. Indeed, in such a region 
$\chi$ remains trapped in the false vacuum for such a long time that  $T$ becomes comparable to the spacetime curvature. 

This effect can even be amplified by considering the non-minimal couplings between the scalar fields and the Ricci scalar $R$ with coefficients of zero mass dimension. Indeed, when $\chi$ is trapped in the false vacuum, the universe has a constant $R$, which is proportional to the square of the Hubble rate of the corresponding de Sitter space. Then the effective mass of $\chi$ is modified by $R$ through a parameter related to those non-minimal couplings. In this paper we therefore include such couplings and explore their effect in a model-independent way\footnote{Refs.~\cite{Salvio:2023blb,Cutrona:2026mxm,Kim:2026zna} considered the effect of a non-minimal coupling on a FOPT within specific models featuring RSB.}. 

A model-independent approach is clearly desirable because one can then apply the result to any specific model belonging to the above-mentioned scenario. We therefore adopt such an approach in this paper.

False-vacuum decay on curved spacetime at $T=0$  was  studied in a number of papers~\cite{Joti:2017fwe,Rajantie:2017ajw,Markkanen:2018pdo}. Here we do not repeat these analyses, but we rather extend them to the regime where both the spacetime curvature and the temperature play a non-negligible role. 

The false-vacuum decay on curved spacetime and at finite temperature can be induced by appropriate finite-$T$ versions~\cite{Linde:1980tt,Linde:1981zj} of the bounce solutions studied by Coleman~\cite{Coleman:1977py}, also in collaboration with De Luccia~\cite{Coleman:1980aw}, and by appropriate finite-$T$ versions of a homogeneous configuration considered by Hawking and Moss~\cite{Hawking:1981fz}. In this work we take into account these configurations at finite $T$ and at finite spacetime curvature to estimate the false-vacuum decay rate through semi-classical methods in the physically relevant parameter space of the RSB scenario. 

We remark that here we are interested in the case in which the symmetry-breaking scale, the value of $\chi$ at the true minimum, is much smaller than the Planck scale for two reasons. One is simply phenomenological: we are not able to probe the Planck scale observationally. The other reason is the requirement that the effective field theory of gravity on which we want to rely be valid. 

Once the role of gravity in supercooled PTs of this kind is understood it is then interesting to investigate whether the corresponding effects can be detected. For this reason, another purpose of this paper is to consider the corresponding predictions for the spectrum of GWs and to compare it with the sensitivities of current and proposed GW detectors. To this end, we need, among other things, to compute quantities of the FOPT such as the nucleation temperature (the temperature at which the true-vacuum bubbles are no longer washed out by the exponential expansion of the universe) and the duration of the FOPT, which we evaluate in this work. We also investigate the possibility to interpret the background of GWs detected by the PTAs as the effect of FOPTs in the RSB scenario. Previous works have proposed such an interpretation, but without fully taking gravity into account in the PT process~\cite{Fujikura:2023lkn,Gouttenoire:2023bqy,Salvio:2023ext,Salvio:2023blb,Puchades-Ibanez:2026ksh}; here we provide a model-independent treatment including the gravitational effects.

Finally, in this paper we also address the following question: how  many e-folds elapse during the exponential expansion that the universe undergoes when $\chi$ is trapped in the false vacuum? This is an interesting question as a prolonged phase of exponential expansion can have an important impact on cosmology. Indeed, the total number of e-folds of the universe's expansion is bounded from above (see e.g.~\cite{Liddle:2003as,WeinbergCosmo}).

The paper is organized as follows. In Sec.~\ref{General structure of the theory} we discuss the general structure of the RSB scenario in both the matter and gravity sectors. Sec.~\ref{Supercooled phase transitions} is devoted to the study of the role of gravity in supercooled PTs through a model-independent approach in the RSB scenario. There we  study thermal versions of the Coleman bounce and the Hawking-Moss configuration, and we extract the nucleation temperature and the duration of the FOPT when they exist. We study the predictions of the RSB scenario for the GW spectra and their detectability in Sec.~\ref{Gravitational-wave signals and detectability}; there we also study the possibility to interpret the background of GWs detected by the PTAs as the effect of those predicted GW spectra. In Sec.~\ref{Number of e-folds} we study the number of e-folds that elapse when $\chi$ is trapped in the false vacuum. Our conclusions are offered in Sec.~\ref{Conclusions}.

\section{General structure of the theory}\label{General structure of the theory}

Let us start describing the RSB matter   sector. We consider the most general no-scale  matter Lagrangian\footnote{We use the mostly minus convention for the metric signature.}:
\begin{equation}
    \label{eq:noscaleL}
    \Lag^{\rm ns}_{\rm matter}\equiv-\frac{1}{4}F_{\mu \nu}^MF^{\mu \nu}_M+ \frac{1}{2}(D_\mu \phi)_a (D^\mu \phi)_a + \bar{\psi}_ji\slashed{D}\psi_j - \frac{1}{2}(Y_{ij}^a \psi_i\psi_j\phi_a+ \text{h.c.})-V_{\rm ns}(\phi).
\end{equation}
We allow for generic numbers 
of real scalars $\phi_a$,   Weyl fermions $\psi_j$ and vectors $V^M_\mu$ (with field strength $F_{\mu\nu}^M$), respectively. The gauge fields $V^M_\mu$ are used to build the covariant derivatives \begin{equation}
    \label{eq:covD}
D_\mu\phi_a=\partial_\mu\phi_a+i\theta_{ab}^MV_\mu^M\phi_b,\hspace{1cm} D_\mu\psi_j=\partial_\mu\psi_j+it_{jk}^MV_\mu^M\psi_k
\end{equation}
($\theta^M$ and $t^M$ are the generators of the gauge group in the scalar and fermion representations, respectively) and $\Lag^{\rm ns}_{\rm matter}$ is invariant under the gauge group. We include the gauge couplings in the generators. Since all scalars are real, the Hermitian matrices $\theta^M$ are all antisymmetric and purely imaginary. 
The $Y^a_{ij}$  are the Yukawa couplings, which are symmetric in $i\leftrightarrow j$,  
\begin{equation}
    \label{eq:noscaleV}
    V_{\rm ns}(\phi) = \frac{\lambda_{abcd}}{4!}\phi_a\phi_b\phi_c\phi_d
\end{equation}
is the no-scale scalar potential and the $\lambda_{abcd}$ are the quartic couplings.  

In the RSB mechanism the scalar potential develops a flat direction at a certain value $\tilde \mu$ of the  renormalization scale $\mu$. Such direction is parametrized by a scalar field $\chi$  as $\phi_a=\chi\nu_a$, where $\nu$ is a unit vector, $\nu_a\nu_a=1$, in the scalar-field space. We call $\chi$ the flat-direction field. Along $\nu$ the RG-improved potential $V$ is
\begin{equation}
   V(\chi)=\frac{\lambda_\chi(\mu)}{4}\chi^4,\hspace{1cm}
    \lambda_\chi(\mu)\equiv\frac{1}{3!}\lambda_{abcd}(\mu) \nu_a\nu_b\nu_c\nu_d.
\end{equation}
The requirement that the flat direction occurs at $\mu=\tilde\mu$  leads to the condition $\lambda_\chi(\tilde \mu)=0$.
The one-loop quantum effective potential renormalized at $\mu=\tilde\mu$ is  then
\begin{equation}
\label{eq:1loopP}
    V_q(\chi)= \frac{\bar{\beta}}{4}\left(\log\frac{\chi}{\chi_0}-\frac{1}{4}\right)\chi^4, \qquad   \bar{\beta}\equiv \left[\mu \frac{d\lambda_\chi}{d\mu}\right]_{\mu=\tilde{\mu}}
\end{equation}
and $\chi_0$ is the scale introduced via dimensional transmutation by the RSB mechanism. We require $\bar \beta>0$ to ensure that $\chi_0$ is not only a stationary point but also the absolute minimum of $V_q$. The flat-direction field $\chi$ radiatively obtains a mass $m_\chi= \sqrt{\bar{\beta}}\chi_0$. 

Let us turn now to the gravity sector. The corresponding Lagrangian density will include generic no-scale non-minimal couplings between the scalars $\phi_a$ in the RSB sector and\footnote{Our conventions for the Riemann tensor $R_{\mu\nu\,\,\, \sigma}^{\quad \rho}$, the Ricci tensor $R_{\mu\nu}$ 
and $R$ are, respectively, $$R_{\mu\nu\,\,\, \sigma}^{\quad \rho} \equiv \partial_{\mu} \Gamma_{\nu \, \sigma}^{\,\rho}- \partial_{\nu} \Gamma_{\mu \, \sigma}^{\,\rho} +  \Gamma_{\mu \, \tau}^{\,\rho}\Gamma_{\nu \,\sigma }^{\,\tau}- \Gamma_{\nu \, \tau}^{\,\rho}\Gamma_{\mu \,\sigma }^{\,\tau}, \quad R_{\mu\nu} \equiv R_{\rho\mu\,\,\, \nu}^{\quad \rho},\quad R\equiv g^{\mu\nu}R_{\mu\nu},$$
where $\Gamma_{\mu \, \sigma}^{\,\rho}$ is the Levi-Civita connection.} $R$, parameterized by dimensionless coefficients $\xi_{ab}$:

 \be \Lag_{\rm gravity} = -\frac{\bp^2}{2} R -\frac12\xi_{ab} \phi_a\phi_b R  - \Lambda_0, \label{Lgravity}\ee  
 where $\bp$ is the reduced Planck mass and $\Lambda_0$ is a contribution to the cosmological constant. The full cosmological constant also receives a contribution from the minimum of the quantum effective potential and at the one-loop level reads
\be  \Lambda = \Lambda_0-\frac1{16}\bar\beta\chi_0^4. \ee 
The term $\Lambda_0$ can be chosen in a way that $\Lambda$ matches the observed value of the dark energy. It is worth mentioning that the $\xi_{ab}$ are needed for the renormalization of the theory on a curved background even if one only considers quantum corrections due to matter fields~\cite{BirrellDavies}.

Note that, because of the non-minimal couplings in~\eqref{Lgravity}, the effective Lagrangian renormalization-group-improved along the flat direction $\nu$ contains a non-minimal coupling of the form
%\mathscr{L}_{\rm eff} \supset
\be  -\frac12 \xi(\mu) \chi^2 R, \qquad \xi(\mu) \equiv \xi_{ab} \nu_a\nu_b. \ee
RSB occurs when $\mu=\tilde\mu$ so we will take into account the non-minimal coupling $\xi(\tilde\mu)$, which we simply call $\xi$ from now on.

In~\eqref{Lgravity} we inserted explicit mass scales, $\bp$ and $\Lambda_0$. It would also be possible to consider a no-scale gravitational Lagrangian and generate $\bp$ and $\Lambda_0$ radiatively~\cite{Salvio:2014soa,Kannike:2015apa,Salvio:2017qkx,Salvio:2017xul,Salvio:2019wcp,Salvio:2024joi,Salvio:2020axm}. However, this fact is not essential to the purpose of the present paper and for the sake of generality we do not specify the origin of $\bp$ and $\Lambda_0$. 
%Here we do not attempt to construct a fully classically-scale-invariant theory because, as we have already pointed out, the RSB sector can also be 

Also, the full theory can include a matter sector that is not classically scale invariant. This is the case when the RSB sector is a ``dark" sector weakly coupled to the SM. In general the full action is  then 
\be S = \int d^4x\sqrt{|\det \hat g|}(\Lag_{\rm matter} +\Lag_{\rm gravity}),  \ee
where $\det \hat g$ is the determinant of the metric and $\Lag_{\rm matter}$ includes both $\Lag^{\rm ns}_{\rm matter}$ and the contribution of the possible matter sector that is not classically scale invariant, if there is one.

In the absence of other ingredients able to UV complete gravity this theory must be considered an effective field theory valid only up to energies below  a cutoff $\Lambda_J$, which cannot be larger than the effective reduced Planck mass,  $\sqrt{\bp^2+\xi \chi^2}$, where here $\chi$ is a generic background value of the flat-direction field: this is  the intrinsic cutoff due to the fact that  gravity is not renormalizable. A large value of $\xi$ can considerably lower $\Lambda_J$~\cite{BurgessLeeTrott,BarbonEspinosa,Burgess:2010zq,Hertzberg,Burgess:2014lza,BezrukovMagnin}. Ref.~\cite{BezrukovMagnin} showed in particular that the cutoff depends on the background field value of the flat-direction field and reads 
\be \Lambda_J (\chi) = \min\Bigg(
\frac{\bp^2+\xi\chi^2+6\xi^2\chi^2}{\xi\sqrt{\bp^2+\xi\chi^2}}, \sqrt{\bp^2+\xi \chi^2}\Bigg).
%\qquad
%\frac{\Lambda_J(\chi_0)}{\chi_0}\ge2\sqrt6 .
\label{eq:LambdaJ}
\ee 
%A model-independent sufficient condition for the validity of a perturbative description of the EFT is that $\chi_0\ll \Lambda_J(\chi_0)$. Indeed, generically a non-zero value of $\chi_0$ can generate particle masses of order $\chi_0$, but not much larger than $\chi_0$ in order for perturbation theory to hold. We will therefore compute  $\Lambda_J\equiv\Lambda_J(\chi_0)$ in this paper.

\section{Supercooled phase transitions}\label{Supercooled phase transitions}

To study the PT we consider the theory at finite temperature $T$. Refs.~\cite{Witten:1980ez,Salvio:2023qgb,Salvio:2023ext,Banerjee:2024cwv,Salvio:2024upo,Salvio:2026bco} showed that the symmetry breaking that $\chi_0$ induces (if any) is associated with a supercooled first-order PT, at least in Einstein gravity with minimal coupling to matter. Supercooling means $T\ll \chi_0$  for the values of $T$ of interest in the PT.  The presence of a non-minimal coupling, $\xi$, can change this result and lead to a parameter space where supercooling does not occur. This can happen when $\xi$ is sufficiently large and positive. Indeed, when $\chi$ is trapped in the symmetry-preserving vacuum, $\chi=0$, the universe inflates so that the effective squared mass of $\chi$ receives a nearly-constant contribution $\xi R$, which is negative for $\xi>0$. 
In this work we are interested in supercooled PTs for the reasons explained in the introduction. Therefore, while we also consider large values of $\xi$, when $\xi>0$, we do not take $\xi$ so large that the PT is not supercooled.

In supercooled phase transitions with RSB the effective potential $V_{\rm eff}$ at finite $T$ is well described by the one-loop approximation (if the usual zero-temperature perturbativity conditions hold)~\cite{Salvio:2023qgb,Salvio:2023ext,Salvio:2026bco} and in the Landau gauge reads
\be V_{\rm eff}(\chi,T) = V_q(\chi) +\frac{T^4}{2\pi^2}\left(\sum_b n_b J_B(m_b^2(\chi)/T^2)-2\sum_f J_F(m_f^2(\chi)/T^2)\right)+\Lambda_0,  \label{VeffSumm}  \ee
where the $m_b$ and $m_f$ are the background-dependent bosonic and fermionic masses, respectively, the sum over $b$ runs over all bosonic degrees of freedom ($n_b=1$ for a real scalar and $n_b=3$ for a real vector), while the  sum over $f$   runs over the fermion degrees of freedom (the factor of  2 is there because we work with Weyl spinors). Also,  
the thermal functions $J_B$ and $J_F$ are 
\bea \, \hspace{-1cm}J_B(x)\equiv \int_0^\infty dp\, p^2 \log\left(1-e^{-\sqrt{p^2+x}}\right)&=&-\frac{\pi^4}{45}+\frac{\pi^2}{12} x -\frac{\pi}{6} x^{3/2} -\frac{x^2}{32} \log\left(\frac{x}{a_B}\right) + O(x^3), \label{JBdef}\\
\hspace{-1cm}J_F(x)\equiv \int_0^\infty dp\, p^2 \log\left(1+e^{-\sqrt{p^2+x}}\right)&=& \frac{7\pi^4}{360}-\frac{\pi^2}{24} x -\frac{x^2}{32} \log\left(\frac{x}{a_F}\right) + O(x^3) \label{JFdef},\eea
 where $a_B = 16\pi^2 \exp(3/2-2\gamma_E)$, $a_F = \pi^2 \exp(3/2-2\gamma_E)$ and  $\gamma_E$ is the Euler-Mascheroni constant.
 
 Note that, as a consequence of supercooling, the derivative and curvature  loop corrections to the effective action can be neglected: they are suppressed by zero-temperature loop factors as $T$ is very small.
 
 \subsection{Supercool expansion and improved supercool expansion}
Refs.~\cite{Salvio:2023qgb,Salvio:2023ext}, working within the framework of minimally coupled matter, found that the phase transitions in no-scale theories with RSB  can be described in a model-independent way and in terms of a few parameters if the following quantity is small, or at least not large,
\be \epsilon\equiv  \frac{g^4}{6\bar\beta \log\frac{\chi_0}{T}},
 \label{CondConv}\ee
 for the values of $T$ of interest in the PT. The parameter $\epsilon$ measures the amount of supercooling: the smaller $\epsilon$, the larger the supercooling, because of the $\log(\chi_0/T)$ factor in the denominator.   Here 
  \be g^2 \equiv \sum_b n_bm_b^2(\chi)/\chi^2+\sum_f m^2_f(\chi)/\chi^2, 
\label{M2g2def}\ee
where $g$ is a  real, non-negative and $\chi$-independent quantity. Since $\bar\beta$ in~\eqref{eq:1loopP} is a beta function, we expect  that the parameter 
\be Y\equiv\frac{\bar\beta(4\pi)^2}{g^4} \label{Ydef}\ee
is at most of order one in the perturbative regime.
We now show that a non-minimal coupling changes the quantity in~\eqref{CondConv}, namely it changes the parameter whose smallness  allows us to describe the PTs in no-scale theories with RSB  in a model-independent way.

After supercooling, but before the false-vacuum decay has taken place, the universe inflates with Hubble rate
\be H_I = \frac{\sqrt{\bar\beta} \chi_0^2}{4\sqrt{3}\bp}.\label{Hif}\ee
We refer to this stage in the history of the universe as thermal inflation as $V_{\rm eff}$ depends on $T$.
The false-vacuum decay rate depends on the Euclidean action of some field configurations, which will be discussed in Secs.~\ref{Hawking-Moss solution}-\ref{Sizable Hubble rate}. These Euclidean actions receive the dominant contribution from field values around the barrier.  
%This is true a posteriori for all field configurations of interest, including the Hawking-Moss (HM) which sits at the top of the barrier. For the other field configurations the reason is similar to that in \cite{Salvio:2023qgb} and \cite{Salvio:2023ext} because, as we have seen, the relevant solutions are the bounces with R_b << \pi/H_I and the HM ones (already discussed). For the bounces with R_b << \pi/H_I, the effect of gravity can be completely encoded in m_{\rm eff} and therefore you can apply the reasons in \cite{Salvio:2023qgb} for the supercool expansion; in the improved supercool expansion I checked that the Euclidean action receives the dominant contribution from field values around the barrier using the explicit computation of the bounces in CW-GW.nb (see also CWgravity).
Therefore, 
we first estimate the barrier size, which we can define as the field value $\chi_b$ at which 
%$V_{\rm eff}$ equals its value at the false vacuum $\chi=0$:
\be \bar V_{\rm eff}(\chi_b,T)  = 0, \label{chibCond}\ee 
where 
\be \bar V_{\rm eff}(\chi,T)\equiv V_{\rm eff}(\chi,T)-6\xi H_I^2\chi^2-V_{\rm eff}(0,T).\ee
The second term in $\bar V_{\rm eff}$ is the effect of the non-minimal coupling. Using the expression of $V_q$ in~\eqref{eq:1loopP} and defining 
\be J_T(\chi^2/T^2) \equiv \sum_b n_b J_B(m_b^2(\chi)/T^2)-2\sum_f J_F(m_f^2(\chi)/T^2),\ee
the condition in~\eqref{chibCond} reads
\be \bar V_{\rm eff}(\chi_b,T)= \frac{\bar{\beta}}{4}\left(\log\frac{\chi_b}{\chi_0}-\frac{1}{4}\right)\chi_b^4 +\frac{T^4}{2\pi^2} (J_T(\chi_b^2/T^2)-J_T(0))- 6\xi H_I^2 \chi_b^2 = 0. \label{chibCond2}   \ee 

 The model-independent description in terms of a few parameters of Refs.~\cite{Salvio:2023qgb,Salvio:2023ext} keeps holding as long as $\chi_b/T$ is small enough so that the first few terms in the small-$x$ expansions in~\eqref{JBdef} and~\eqref{JFdef} already provide a good approximation. 
 %Supercooling implies indeed that \chi_b is small unless \xi<0 and large, but in that case we would have a regime in which the temperature is completely negligible and therefore one can use the results of Joti et al 
 For a sufficiently small $\chi_b/T$ the condition in~\eqref{chibCond2} leads to
 \be \frac{\chi_b^2}{T^2} = \frac{2 J_T'(0)}{\pi^2 \bar\beta\left( \log\frac{\chi_0}{\chi_b} +\frac14\right)} -\frac{24\xi H_I^2}{\bar\beta \, T^2\left( \log\frac{\chi_0}{\chi_b} +\frac14\right)} \lesssim \frac{g^2}{6\bar\beta X} -\frac{24\xi H_I^2}{\bar\beta \, T^2 X}, \label{chibTbound}\ee 
where we have defined 
\be X\equiv \log\frac{\chi_0}{T} \label{xDef}\ee
 and in the last step we used the fact that $\log\frac{\chi_0}{\chi_b} +\frac14  \gtrsim X$ for a sufficiently small $\chi_b/T$ and\footnote{As usual, a prime on a function denotes the derivative with respect to its argument.} 
 %the derivative of $J_T$ with respect to $\chi^2/T^2$
  $J_T'(0) = \pi^2g^2/12$. Noting now that $\chi/T$ appears in~\eqref{VeffSumm} always multiplied by a coupling constant, the  condition that $\chi_b/T$ be small enough (so that the first few terms in the small-$x$ expansions in~\eqref{JBdef} and~\eqref{JFdef} already provide a good approximation)  leads to the requirement that 
\be \epsilon_g \equiv \epsilon - \frac{24g^2\xi H_I^2}{\bar\beta \, T^2 X} \label{epsilong}\ee
be small enough. So the gravitational non-minimal coupling $\xi$ deforms the quantity that sets the validity of the model-independent treatment\footnote{That deformation was not taken into account in Ref.~\cite{Cutrona:2026mxm}, however, the deformation leads to negligibly small corrections in all cases considered in Ref.~\cite{Cutrona:2026mxm}.}.  The quantity in~\eqref{epsilong} can equivalently be written as follows
 \be \epsilon_g=\epsilon \left(1- s_\xi e^{2(X-X_m)}\right), \label{epsilong2} \ee 
 where $s_\xi$ is the sign of $\xi$ and 
 \be X_m\equiv \frac12 \log\left(\frac{g^2 \bp^2}{3|\xi| \bar\beta \chi_0^2}\right). \label{XmDef} \ee
 
 When $\epsilon_g\ll1$ and in the small-$x$ expansions of Eqs.~\eqref{JBdef} and~\eqref{JFdef} the $x^{3/2}$ terms give a small contribution compared with the $x$ terms, one can start by considering only the leading $x$ terms   and then treat all the remaining terms as small perturbations. At the leading order (LO) in this perturbative treatment (which we refer to as the supercool expansion), we have 
 \be V_{\rm eff}(\chi,T)\approx \frac{m^2}{2} \chi^2 - \frac{\lambda}{4} \chi^4 + V_{\rm eff}(0,T)\label{Vapp}, \ee
  where 
 \be m \equiv \frac{g \,T}{\sqrt{12}}, \qquad \lambda \equiv \bar\beta X, \qquad V_{\rm eff}(0,T) = \Lambda_0+ \frac{T^4}{2\pi^2} J_T(0) .   \label{Defml}\ee
 Equivalently, for $\bar V_{\rm eff}$ we obtain the approximation
 \be \bar V_{\rm eff}(\chi,T)\approx \frac{m_{\rm eff}^2}{2} \chi^2 - \frac{\lambda}{4} \chi^4 \label{barVapp}, \ee
 where 
 \be \boxed{m_{\rm eff}^2 \equiv \frac{g^2 T^2}{12}-12\xi H_I^2.}  \label{meffDef} \ee 
Note that $m_{\rm eff}^2$ includes the effect of the non-minimal coupling. 
The quartic term in~\eqref{barVapp} has coefficient $- \lambda/4$ because the smallness of $\chi_b/T$ implied by~\eqref{chibTbound} tells us 
 \be \log\frac{\chi_b}{\chi_0} -\frac14= \log\frac{\chi_b}{T}-\frac14-X \approx -X \label{logSplit}\ee 
 because the right-hand side of~\eqref{chibTbound} suppresses $\chi_b/T$ only logarithmically in $\chi_0/T$. 
 When the supercool expansion is performed to higher orders, including not only the $x^{3/2}$ term but also  the $x^2\log x$ and $x^2$ terms in the small-$x$ expansions of~(\ref{JBdef})-(\ref{JFdef}) to obtain higher accuracy, generically one should also take into account the $\chi_b$-dependent logarithm in~(\ref{logSplit}) and the approximation in~\eqref{logSplit} is not  valid to obtain such higher accuracy. 
 
 However, the supercool expansion at leading order (LO), in which $\bar V_{\rm eff}(\chi,T)$ is approximated as in~\eqref{barVapp}, is accurate enough to compare the theoretical predictions with GW observations if $\epsilon_g\ll1$ and the $x^{3/2}$ terms give a small contribution compared with the $x$ terms in Eq.~\eqref{JBdef}. In fact this remains true even if one only has $\epsilon_g\sim1$, as long as there are many degrees of freedom  with dominant couplings (all of the same order of magnitude) to $\chi$. These results were shown in~\cite{Salvio:2023ext,Salvio:2026bco} considering only minimal couplings of the matter fields to gravity. However, we checked that the same remains true even for $\xi\neq0$.

 If $\epsilon_g\sim1$, but there are not many degrees of freedom  with dominant couplings to $\chi$, the supercool expansion breaks down because the term of order $x^{3/2}$ in the small-$x$ expansion of~(\ref{JBdef}) can be comparable to the one of order $x$. In this case one must substitute the supercool expansion with an improved supercool expansion, for which the LO approximation of $ V_{\rm eff}$ includes a cubic term in $\chi$:
 \be V_{\rm eff}(\chi,T) \approx \frac{m^2}{2} \chi^2-\frac{k}{3}\chi^3-\frac{\lambda}{4} \chi^4+V_{\rm eff}(0,T), \label{Vnlo}\ee
 where 
 \be k\equiv \frac{\tilde g^3 T}{4\pi}, \qquad \tilde g^3 \equiv \sum_b n_bm_b^3(\chi)/\chi^3.\label{kdef}\ee
$\tilde g$ is a  real, non-negative and $\chi$-independent parameter, satisfying $\tilde g\leq g$.
Equivalently,
\be \bar V_{\rm eff}(\chi,T) \approx \frac{m^2_{\rm eff}}{2} \chi^2-\frac{k}{3}\chi^3-\frac{\lambda}{4} \chi^4. \label{barVnlo}\ee
In the improved supercool expansion one treats all terms beyond the order $x^{3/2}$ in the small-$x$ expansions of~(\ref{JBdef})-(\ref{JFdef}) as small perturbations. Refs.~\cite{Salvio:2023ext,Salvio:2026bco} showed that the improved supercool expansion at LO is accurate enough to compare the theoretical predictions with GW observations if $\epsilon_g$ is at most of order one for $\xi =0$, but we checked that the same result remains true for $\xi\neq0$.
 
 The effect of $\xi$ in $\bar V_{\rm eff}$ is to modify the barrier, Eq.~\eqref{meffDef}. We can equivalently write 
 \be m^2_{\rm eff}= m^2\left(1- s_\xi e^{2(X-X_m)}\right). \label{meffs} \ee 
 If $\xi>0$ (namely $s_\xi=1$) the barrier disappears when the temperature drops to a value such that $X=X_m$, below which the field rolls towards the symmetry-breaking vacuum. If $\xi<0$ (namely $s_\xi=-1$) the barrier never disappears and at the temperature such that $X=X_m$ the thermal contribution to $m^2_{\rm eff}(T)$ simply equals the gravitational contribution.

\subsection{Hawking-Moss solution}\label{Hawking-Moss solution}

The Hawking-Moss (HM) configuration is a de Sitter solution with Hubble rate $H$ of the Euclidean field equations for the metric and $\chi$, with $\chi$ at the spacetime-independent (homogeneous) value $\chi_{\rm top}$ that maximises $V_{\rm eff}-12 H^2\xi\chi^2$ at fixed $H$~\cite{Hawking:1981fz}. In the HM solution $\chi$  and the Hubble rate $H$ therefore satisfy the equations 
\bea &&\frac{d V_{\rm eff}}{d\chi} - 12 H^2 \xi \chi = 0, \label{chiEqHM}\\
&&H^2 = \frac{V_{\rm eff}}{3 \bp^2(1+\xi \chi^2/\bp^2)}.\eea 
The HM configuration exists whenever there is a maximum of $V_{\rm eff}-6 H^2\xi\chi^2$ at fixed $H$.

As we have discussed, we are interested in cases where $T$ plays a non-negligible role,
%this is not to reduce to the Joti et al analysis. Therefore the H contribution to the barrier height is not dominant compared to the thermal contribution
 but the phase transition is supercooled. In these cases we can approximate $H\approx H_I$, which is the value of $H$ at $\chi=0$ in a supercooled universe, given in\footnote{Note that the approximation of quasi-thermodynamic equilibrium in such an expanding universe still holds as long as the rate of processes maintaining thermal equilibrium is large compared to $H_I$; this is the case, unless the couplings are really tiny,  because such rate is of order $\chi_0$ times coupling constants~\cite{Rescigno:2025ong}, and $H_I$ is Planck suppressed compared to $\chi_0$, as clear from~\eqref{Hif}. In our notation $T$ is the temperature of the cosmological plasma, which is distinct from the de Sitter effective temperature, $H_I/(2\pi)$, defined by Gibbons and Hawking~\cite{Gibbons:1977mu}.}~\eqref{Hif}. Therefore, Eq.~\eqref{chiEqHM}, which determines $\chi$ in the HM solution, reduces to
 \be \frac{d V_{\rm eff}}{d\chi} - 12 H_I^2 \xi \chi = 0. \label{chiEqHMp} \ee 
 For example, in the supercool expansion at LO when $V_{\rm eff}$ is approximated as in~\eqref{Vapp} we find
\be  \chi_{\rm top}^2 = \frac{m_{\rm eff}^2}{\lambda}, \ee
with $m_{\rm eff}^2$ and $\lambda$ defined in~\eqref{meffDef} and~\eqref{Defml}. On the other hand, in the improved supercool expansion at LO when $V_{\rm eff}$ is approximated as in~\eqref{Vnlo} we find
\be  \chi_{\rm top} = \frac{\sqrt{k^2+4m_{\rm eff}^2\lambda}-k}{2\lambda}, \label{chitop}\ee 
with $k$ defined in~\eqref{kdef}. Note that we can obtain the result in the supercool expansion at LO setting $k=0$ in the improved version.

In the cases we are interested in here
%as stated above we are interested in cases where $T$ plays a non-negligible role, but the phase transition is supercooled. 
we have
\be V_{\rm eff}(\chi_{\rm top},T)-V_{\rm eff}(0,T)\ll V_{\rm eff}(0,T). \label{deltaVc} \ee
%This is true in the cases we are interested in: for \xi>0 the barrier in \bar V_{eff} is smaller than that in V_{eff} so evaluating V_{\rm eff} at \chi_{top} gives a value closer to V_{\rm eff}(0,T) than evaluating V_{\rm eff} at its maximum; for \xi<0 we don't take |\xi|so large that the temperature contribution to m_{eff} is negligible to the  H contribution to m_{eff} (in order not to reduce to the Joti et al analysis)
The condition in~\eqref{deltaVc} tells us that the  difference $S_{\rm HM}$ between the Euclidean action of the HM solution and that of the de Sitter solution with $\chi=0$ is well approximated by 
\begin{equation}
S_{\rm HM}\approx
\frac{8\pi^2}{3H_I^4}\,\bar V_{\rm eff}(\chi_{\rm top},T).
\label{eq:SHM_geom}
\end{equation}
%
%\qquad
%\Delta V_{\rm top}\equiv V_{\rm eff}(\chi_{\rm top},T)-V_{\rm eff}(0)
Here $S_{\rm HM}$ plays an important role because we will estimate the HM contribution to the probability of false-vacuum decay through the semi-classical exponential  $\exp(-S_{\rm HM})$.
In the improved supercool expansion at LO, Eq.~\eqref{chiEqHMp} leads to
\begin{equation}
\lambda\chi_{\rm top}^2+k\chi_{\rm top}=m_{\rm eff}^2,
\label{eq:saddle_imp}
\end{equation}
and therefore we obtain
\be \bar V_{\rm eff}(\chi_{\rm top},T)=\frac{k}{6}\chi^3_{\rm top}+\frac{\lambda}{4}\chi^4_{\rm top}.  \label{Vbctop}\ee
Again we can obtain the result in the supercool expansion at LO setting $k=0$ in the improved version.
Using~\eqref{Vbctop} in~\eqref{eq:SHM_geom} we obtain $S_{\rm HM}$ and we can estimate the HM contribution to the probability of false-vacuum decay.

As remarked by Hawking and Moss~\cite{Hawking:1981fz}, the HM solution, featuring a homogeneous $\chi$ at the top of $V_{\rm eff}-6 H^2\xi\chi^2$, is expected, in Lorentzian signature, to evolve according to the classical field equations, with
$\chi$ rolling  from that maximum towards the symmetry-breaking vacuum, remaining spatially homogeneous. In this case no bubbles  form and no GWs can be produced through bubble collisions.

\subsection{Solutions with a  non-homogeneous scalar field}\label{Solutions with a  non-homogeneous scalar field}

Bubbles can be represented by solutions with a non-homogeneous $\chi$, the bounces. For $T=0$ solutions of this type were found by Coleman (without gravity)~\cite{Coleman:1977py} and later by Coleman and de Luccia (with gravity)~\cite{Coleman:1980aw}. 
Linde~\cite{Linde:1980tt,Linde:1981zj} showed how $T$ %(much higher than the size of the bubble $R_b$) 
changes the Coleman bounce in~\cite{Coleman:1977py}. Here we want to include both the temperature and gravity.  As discussed in the introduction, we are interested in cases where $T$ plays a non-negligible role,
 but the phase transition is supercooled. In these cases, the backreaction of a bounce solution to gravity is negligible, 
 %this is because the  Euclidean actions of the bounces receive the dominant contribution from field values around the barrier, as we have seen. 
% and we can approximate $H\approx H_I$, given in~\eqref{Hif}. 
so we are interested  in studying bounces on a fixed de Sitter background with Hubble rate $H_I$.

 Bounces of this type only exist if $H_I$ is smaller than a critical value $H_{\rm cr}$. Working at zero temperature, Refs.\cite{Hawking:1981fz,Joti:2017fwe} pointed out that this critical value satisfies 
 \be  H_{\rm cr}^2= -\frac14 \bar V_{\rm eff}^{(2)}(\chi_{\rm top},T), \label{Hcond}\ee 
 where $\bar V_{\rm eff}^{(2)}$ is the second derivative of $\bar V_{\rm eff}$ with respect to its first argument, $\chi$. 
 A finite value of $T$, besides modifying the effective potential, also requires the imaginary time of Euclidean spacetime to be periodic with period $1/T$. This is an extra boundary condition to be imposed on the bounces. As a result, the condition in~\eqref{Hcond} remains a necessary condition for the existence of bounces of the type we are interested in. For example, in the supercool expansion at LO when $\bar V_{\rm eff}$ is approximated as in~\eqref{barVapp} we find that $H_{\rm cr}$ satisfies
 \be  H_{\rm cr}^2(1+6\xi) = \frac{m^2}{2}, \label{HcrSE}\ee
%with $m_{\rm eff}$ and $\lambda$ defined in~\eqref{meffDef} and~\eqref{Defml}. 
On the other hand, in the improved supercool expansion at LO when $\bar V_{\rm eff}$ is approximated as in~\eqref{barVnlo} we find that $H_{\rm cr}$ satisfies
 \be  H_{\rm cr}^2(1+6\xi) = \frac{m^2}{2}+\frac{k^2}{8\lambda} -\frac{k}{8\lambda}\sqrt{k^2+4m^2\lambda -48 \xi \lambda H_{\rm cr}^2}. \label{HcrISE}\ee
For negligibly small values of $\xi$ Eqs.~\eqref{HcrSE} and~\eqref{HcrISE} respectively become
 \be  H_{\rm cr}^2 = \frac{m^2}{2}, \qquad H_{\rm cr}^2=\frac{m^2}{2} +\frac{k^2}{8\lambda} \left(1- \sqrt{1+\frac{4m^2\lambda}{k^2}}\right)< \frac{m^2}{2}. \label{Hcrxi0}\ee
To obtain the inequality on the right we used $m>0$, $\lambda>0$ and $k>0$.
 Therefore, $H_{\rm cr}$ is smaller at $k\neq0$ than at $k=0$. Note that the temperature dependence of $m$, $k$ and $\lambda$ implies a temperature dependence of $H_{\rm cr}$.
 Bounces of the type described at the beginning of this section can only exist when there is a barrier in $\bar V_{\rm eff}$. Within the supercool expansion and the improved supercool expansion this corresponds to the requirement $m_{\rm eff}^2>0$.

The bubbles created are diluted by the expansion of the universe and they cannot collide until $T$ reaches the nucleation temperature $T_n$,  the temperature at which the false-vacuum decay rate per unit volume $\Gamma$  becomes of order  $H_I^4$. The equation $\Gamma=H_I^4$ does not always admit a solution with respect to $T$. Since the barrier in $\bar V_{\rm eff}$ is reduced (enhanced) for $\xi>0$ ($\xi<0$), the parameter space for which there exists a solution of  $\Gamma=H_I^4$ is enhanced (reduced) correspondingly.

An important quantity in the description of a PT  is its inverse duration $\beta$ that, in models with supercooling, is given by~\cite{Caprini:2015zlo,Caprini:2018mtu,vonHarling:2019gme} 
  \be\beta = \left[\frac1{\Gamma}\frac{d\Gamma}{dt}\right]_{t_n}, \label{betaDef}\ee 
  where   $t_n$ is the value of the cosmic time $t$ when $T=T_n$.

We will study the relevant solutions, whose Euclidean actions determine the false-vacuum decay rate (at the semi-classical level),  first assuming $H_I$ to be sufficiently small in Sec.~\ref{Small Hubble rate regime} and then for sizable $H_I$ in Sec.~\ref{Sizable Hubble rate}.

\subsection{Small-bounce regime}\label{Small Hubble rate regime}
Here we consider the small Hubble rate regime in the supercool expansion and in the improved supercool expansion. What we mean by that is  $H_I/\pi\ll m_{\rm eff}$. The quantity $m_{\rm eff}$ is positive  ($\bar V_{\rm eff}$ has to have a barrier). When $H_I$ satisfies this condition the bounces that we discussed in Sec.~\ref{Solutions with a  non-homogeneous scalar field} can only feel the spacetime curvature through $m_{\rm eff}^2$ because the  characteristic bounce size $R_b$ is of order $R_b\sim 1/m_{\rm eff}$, so that  the size of the bounce is small compared to $\pi/H_I$, the characteristic length size of Euclidean de Sitter spacetime.
%Considering  Eq. (25) of Joti et al and expanding for \Sin[Hr]/\pi for small \varepsilon \equiv Hr/\pi I find \varepsilon +c3 \varepsilon^3 + c5 \varepsilon^5 +... where c3, c5, ...  are at most of order one
In this regime we can consistently have a non-trivial effect of the spacetime curvature only if $|\xi|\gg 1/(12\pi^2)$, otherwise the contribution of $H_I^2$ to $m^2_{\rm eff}$ (namely the term $-12\xi H_I^2$ in~\eqref{meffDef}) would be negligible compared to $m^2$ just like $H_I^2/\pi^2$ is negligible compared  to $m_{\rm eff}^2$. 

On the other hand, as discussed in the introduction, we are interested in cases where $T$ plays a non-negligible role, so $-12\xi H_I^2$ can be at most of order $m^2$. In these cases the  characteristic bounce size satisfies
\be \frac1{m_{\rm eff}} \equiv\frac{1}{\sqrt{m^2-12\xi H_I^2}} \gtrsim  \frac{\sqrt{12}}{g T}\gtrsim \frac1{T},  \label{meffmT}\ee 
where the definition of $m^2$ in~\eqref{Defml} was used and in the last step we noted that $g$ cannot be large in order for perturbation theory to work.  This result tells us that  the bounce solutions  are approximately time-independent (see~\cite{Linde:1980tt,Linde:1981zj}) and their action $S$ reads
\be S =\frac{S_3}{T}, \qquad  \quad S_3 \equiv 4\pi \int_0^\infty dr \, r^2\left(\frac12 \chi'^2+\bar V_{\rm eff}(\chi,T)\right), \label{S3f}\ee
where a prime here denotes a derivative with respect to the spatial radius $r$.
%\equiv \sqrt{\vec{x}^{\,2}}
Following Ref.~\cite{Coleman:1977th}, $S_3$ evaluated at the time-independent bounce can be expressed in the following simplified way: 
\be S_3 =  -8\pi \int_0^\infty dr \, r^{2} \bar V_{\rm eff}(\chi,T). \label{bounceSd}\ee

In the supercool expansion at LO (when $\bar V_{\rm eff}$ is approximated as in~\eqref{barVapp}), following~\cite{Brezin-Parisi,Arnold:1991cv,Salvio:2023qgb} we find
\be S_3=c_3 \frac{m_{\rm eff}}{\lambda}, \qquad c_3=18.8973...\, .\label{S3c3}\ee
The result in~\eqref{meffmT} tells us that the dominant energy scale in the problem is $T$, therefore the pre-factor in the semi-classical expression for $\Gamma$ is of order $T^4$:
\be \Gamma \sim T^4  e^{-S_3/T}. \label{GammaI}\ee 
We approximate $T_n$ with the solution of the equation $\Gamma=H_I^4$, which, using~\eqref{GammaI}, reads
\be   \frac{S_3}{T} =  4 \log \left(\frac{T}{H_I}\right).\label{TnEq0}\ee
In the supercool expansion at LO, when~\eqref{S3c3} holds, this equation becomes (multiplying by $X$)
\be 4X^2-cX+a\sqrt{1- s_\xi e^{2(X-X_m)}}=0, \label{TnEqsb}\ee 
where the expression of $m_{\rm eff}$ in~\eqref{meffs} was used and we defined
\be a\equiv \frac{c_3g}{\sqrt{12}\bar\beta}, \quad c\equiv 4\log\frac{\chi_0}{H_I} \equiv 4\log\frac{4\sqrt{3}\bp}{\sqrt{\bar\beta}\,\chi_0}. \label{caDef}\ee
As long as the equation that determines $T_n$ is~\eqref{TnEq0}, setting $T=T_n$ in $S_3$ we obtain a remarkably simple formula:
\begin{equation}
\frac{S_3}{T_n}=c-4X_n=4\delta+2\log\frac{144\,|\xi|}{g^2},
\label{eq:S3T_identity}
\end{equation}
 where $X_n$ is the value of $X$ at $T=T_n$ and
\be \delta\equiv X_m-X_n.  \label{deltaDef}\ee 

Here we are interested in the solution of Eq.~\eqref{TnEqsb} with the smaller $X$ as we imagine to start decreasing the temperature (therefore increasing $X$) from a value of order the critical temperature where Eq.~\eqref{TnEq0} has no solution because of supercooling. 
  As we have seen in  Sec.~\ref{Solutions with a  non-homogeneous scalar field}, the parameter space for which there exists a solution for  $T_n$ is enhanced (reduced) for $\xi>0$ ($\xi<0$). This can be explicitly checked by looking at Eq.~\eqref{TnEqsb}: $\xi>0$ reduces the square-root  term on the left-hand-side of Eq.~\eqref{TnEqsb}, and roots appear where the equation without gravity has none ($c^2<16a$): this is the new parameter space allowed by $\xi>0$, whose elements we call the ``gravitational cells". The parameter space for which there exists a solution for $T_n$ at $\xi=0$ is instead made of ``thermal cells" in our terminology.

 It is interesting to note that Eq.~\eqref{TnEqsb} can be brought into a form that has been studied before in the mathematical literature~\cite{MezoBaricz}. 
Isolating the square root on the left-hand side of Eq.~\eqref{TnEqsb} and squaring,
\begin{equation}
a^2\big[1-s_\xi e^{2(X-X_m)}\big]=\big(cX-4X^2\big)^2
\quad\Longrightarrow\quad
e^{2X}=e^{2X_m}\frac{\big(4X^2-cX+a\big)\big(a+cX-4X^2\big)}{s_\xi a^2 }. \label{1stLambert}
\end{equation}
In the variable $v\equiv-2X$ this equation reads
\begin{equation}
e^{v}\prod_{j=1}^{4}\big(v-t_j\big)=-s_\xi a^2 e^{-2X_m},\qquad
t_{1,2}\equiv\frac{-c\pm\sqrt{c^2+16a}}{4},\quad
t_{3,4}\equiv\frac{-c\pm\sqrt{c^2-16a}}{4}.
\label{eq:lambert}
\end{equation}
The left-hand side is $e^{v}$ times a polynomial, whose (generally multivalued) inverse defines the generalized Lambert function $W$ of Ref.~\cite{MezoBaricz}, here with four upper parameters and no lower one:
\begin{equation}
X=-\tfrac12\,W\big(\begin{smallmatrix}t_1\,t_2\,t_3\,t_4\\ \phantom{-}\end{smallmatrix};-s_\xi a^2 e^{-2X_m}\big). \label{GenLamb}
\end{equation}
The parameters $-t_{3,4}/2$ are the two roots of Eq.~\eqref{TnEqsb} without gravity, $4X^2-cX+a=0$; on the gravitational cells, $c^2<16a$, they become complex. Note that squaring to go from Eq.~\eqref{TnEqsb} to Eq.~\eqref{1stLambert} may introduce spurious roots: the solutions of Eq.~\eqref{TnEqsb} satisfy $cX-4X^2\ge0$, i.e.\ $0\leq X\le c/4=\log(\chi_0/H_I)$, or $H_I\leq T_n\leq \chi_0$. As we have seen, here we are interested in the solution of Eq.~\eqref{TnEqsb} with the smaller $X$.  So we find an analytic expression for $X_n$ by selecting the branch\footnote{There are at most five real branches of $W$, which correspond to the possible five zeros of $e^{v}\prod_{j=1}^{4}\big(v-t_j\big)$ at $v=-\infty$, and $v=t_i$. Additional real branches cannot exist because the derivative of $e^{v}\prod_{j=1}^{4}\big(v-t_j\big)$ with respect to $v$ has at most four finite zeros. Since in neighborhoods of $v=-\infty$, and $v=t_i$ the derivative of $e^{v}\prod_{j=1}^{4}\big(v-t_j\big)$ is not zero we can express the branches of $W$ through series via the Lagrange inversion theorem.}
%https://mathworld.wolfram.com/LagrangeInversionTheorem.html
 of $W$ leading via Eq.~\eqref{GenLamb} to the smaller non-negative $X$ satisfying $X\le c/4$. This analytic procedure is rather involved, so we also produced tables containing the numerical determination of $X_n$, which can be downloaded from~\cite{dataset}.

Using~\eqref{GammaI} and the cosmological relation $dt = -dT/(TH)$, where $H$ is the Hubble rate,  the inverse duration of the PT, defined in~\eqref{betaDef}, is given by
\be \frac{\beta}{H_n} \approx T\frac{d}{dT} \frac{S_3}{T}-4, \label{beetaoH}\ee 
where $H_n\approx H_I$ is the Hubble rate at the nucleation temperature.

In the supercool expansion at LO, when~\eqref{S3c3} holds, we find 
\be \frac{\beta}{H_n} \approx a\frac{1+s_\xi (X_n-1)e^{2 (X_n-X_m)}}{X_n^2 \sqrt{1-s_\xi e^{2 (X_n-X_m)}} }-4, \label{betaH}\ee  
where  we used the expression of $m_{\rm eff}$ in~\eqref{meffs}. To proceed we need $X_n$. The solution for $X_n$ previously discussed admits a simple expression when $\delta$
is very small, namely when $X_n$ is very close to $X_m$.
%A simple analytic approximation of the solution of Eq.~\eqref{TnEqsb} we are interested in can be found if we assume that $\delta$
%is very small, namely we can determine solutions $X_n$ of Eq.~\eqref{TnEqsb} which are very close to $X_m$. 
If $\xi>0$, this occurs when the barrier in $\bar V_{\rm eff}$ has almost disappeared; if $\xi<0$, this occurs when  the thermal and gravitational contributions to $m^2_{\rm eff}(T)$ are roughly equal to each other. We find\footnote{To derive~\eqref{eq:delta_LO} we assumed $c-4X_m>0$. However, this is always true in the regime of validity of our approximations: using~\eqref{XmDef} and  the second expression of $c$ in~\eqref{caDef} we find
\be c-4X_m = 2\log\frac{144\,|\xi|}{g^2}, \ee 
which is always positive as long as $|\xi|\gg 1/(12 \pi^2)$ (which we assume in this section) and $g$ is not so large as to invalidate perturbation theory.}
\bea
\delta&\approx&\frac{X_m^2\,(c-4X_m)^2}{2a^2}\quad(\delta\ll1, \xi>0), \label{eq:delta_LO}\\
\delta&\approx&\frac{2 \left(\sqrt{2} a-c X_m+4 X_m^2\right)}{\sqrt{2} a-2 c+16 X_m}\quad(\delta\ll1, \xi<0).
\label{eq:delta_LOn}
\eea
%These solutions hold  when $X_m$ nearly satisfies X_m^2\,(c-4X_m)^2 \approx 0 and \sqrt{2} a-c X_m+4 X_m^2, respectively. 
Substituting~\eqref{eq:delta_LO}-\eqref{eq:delta_LOn} into~\eqref{betaH}, we find a simple expression of $\beta/H_n$ which is valid for  $\delta\ll1$.
However, here we are interested in values of $\delta$ that are not small for the following reason. For $\xi>0$, a small $\delta$ corresponds to a PT of short duration because a small $\delta$ enhances $\beta$:
\be \frac{\beta}{H_n} \approx \frac{a}{X_m \sqrt{2\delta} }-4\quad(\delta\ll1, \xi>0).\ee  
As 
%well known (see e.g.~\cite{Caprini:2015zlo}), 
 we will discuss in Sec.~\ref{Gravitational-wave signals and detectability}, 
a large $\beta/H_n$ tends to suppress the spectrum of GWs emitted by PTs in the RSB scenario. 
Therefore, values of $\delta$ that are the interesting ones for $\xi>0$. This is not the case for $\xi<0$; and we have 
\be \frac{\beta}{H_n} \approx  \frac{\sqrt{2} a}{X_m^2}-\frac{a}{\sqrt{2} X_m}-4 +\frac{a   \left(3 X_m^2-4 X_m+8\right)\delta}{2 \sqrt{2} X_m^3}\quad(\delta\ll1, \xi<0).\ee

\begin{figure}[t!]
%\centering
\includegraphics[scale=1.0]%[width=\linewidth,height=0.89\textheight,keepaspectratio]
{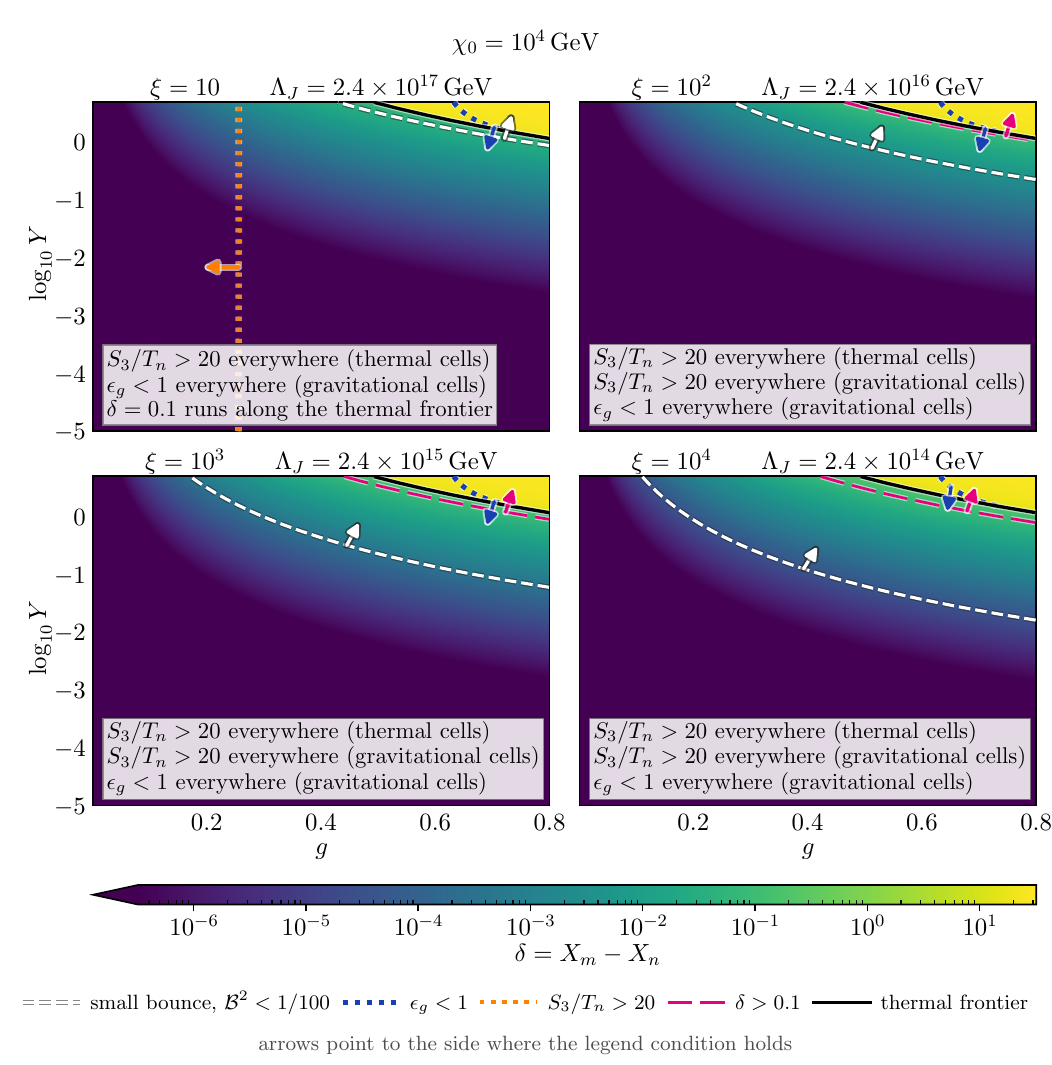}
\caption{\it  Nucleation temperature expressed in terms of $\delta$ (see Eqs.~\eqref{xDef} and~\eqref{deltaDef}) as a function of $g$ and $Y$ (defined in Eqs.~\eqref{M2g2def}-\eqref{Ydef}) in the supercool expansion at LO with gravity for $\xi>0$.  The thermal cells exist above the black line. Arrows point to the side where each condition holds; the conditions that hold everywhere and the conditions that hold nowhere are mentioned in a corner note, here and in the following figures. Here and in the other plots we find a value of $\Lambda_J$ that is field independent with very good accuracy.}
\label{fig:LO_maps}
\end{figure}

In Fig.~\ref{fig:LO_maps} we show the nucleation temperature varying $g$ and $Y$ in the supercool expansion at LO (when $V_{\rm eff}$ is approximated as in~\eqref{Vapp}) with gravity ($\xi>0$) for a value of $\chi_0$ relevant for radiative electroweak symmetry breaking. In the plot we defined $\mathcal B\equiv  [H_I/(\pi m_{\rm eff})]_{T_n}$,
which, as we have seen, has to be small in order for the small-bounce approximation to be valid. We give values of $\Lambda_J$, $S_3/T_n$, $\epsilon_g$ and $\mathcal B$
to understand the validity of our approximations. Also we show where $\delta=0.1$ to understand where $\delta$ is smaller and the GW spectrum is more suppressed.  

\begin{figure}[t!]
\includegraphics[scale=1.0]{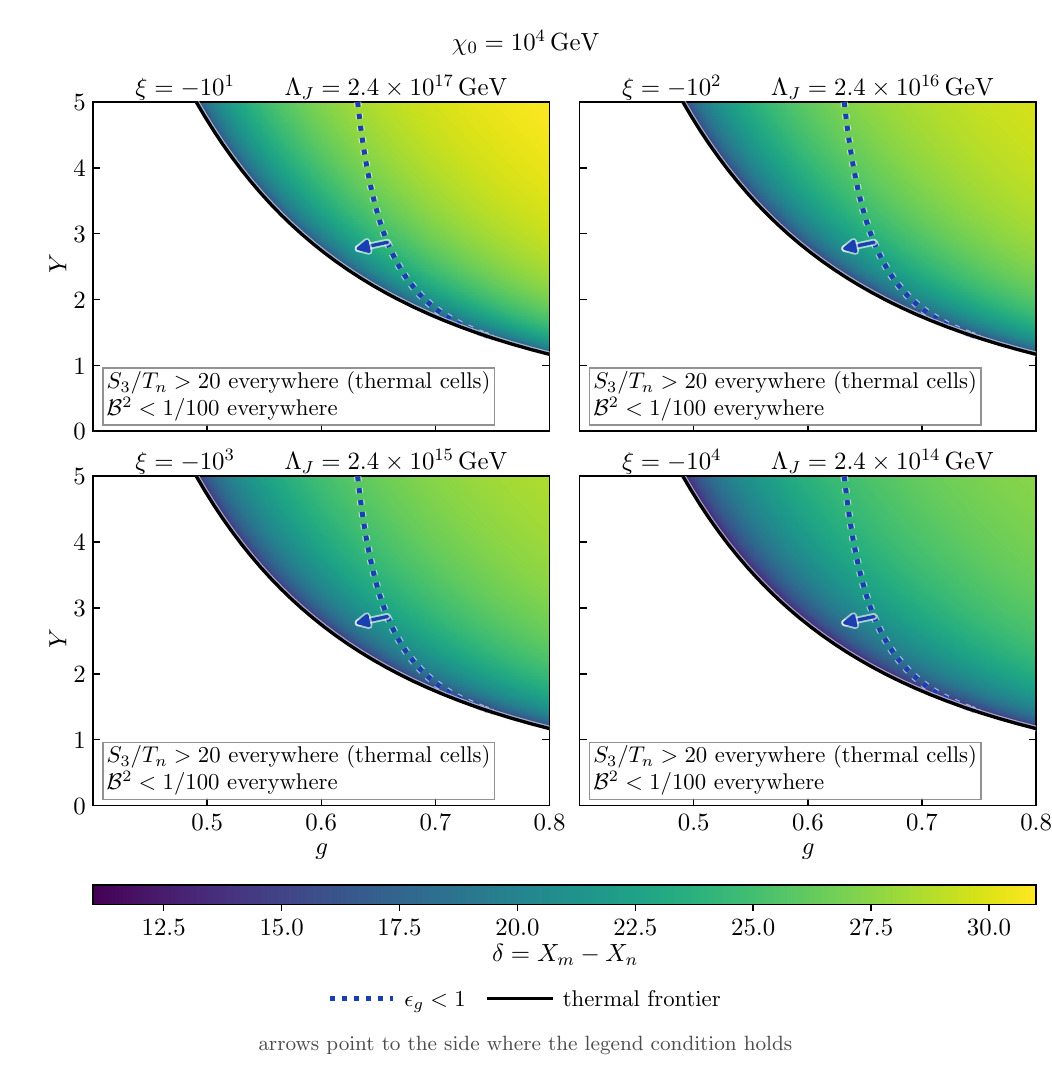}
\caption{\it Nucleation temperature expressed in terms of $\delta$  in the supercool expansion at LO with gravity for $\xi<0$. All cells are thermal (there are no gravitational cells). 
}
\label{fig:LO_neg}
\end{figure}

Fig.~\ref{fig:LO_neg} shows the nucleation temperature in the same setup, but  for $\xi<0$. 
In Fig.~\ref{fig:LO_neg} we do not display the region where $\delta$ is not small because a small $\delta$ is not linked to a suppressed  GW spectrum for $\xi<0$.
 In this case gravity does not lead to further solutions for the nucleation temperature. This result agrees with the fact that $\xi<0$ enhances the barrier in $\bar V_{eff}$.  We also find that where there is a solution the effect of $\xi$ is negligible (all cells are thermal) even taking $|\xi|$ up to $10^4$.

\begin{figure}[t]
\begin{center}
  \includegraphics[scale=0.51]{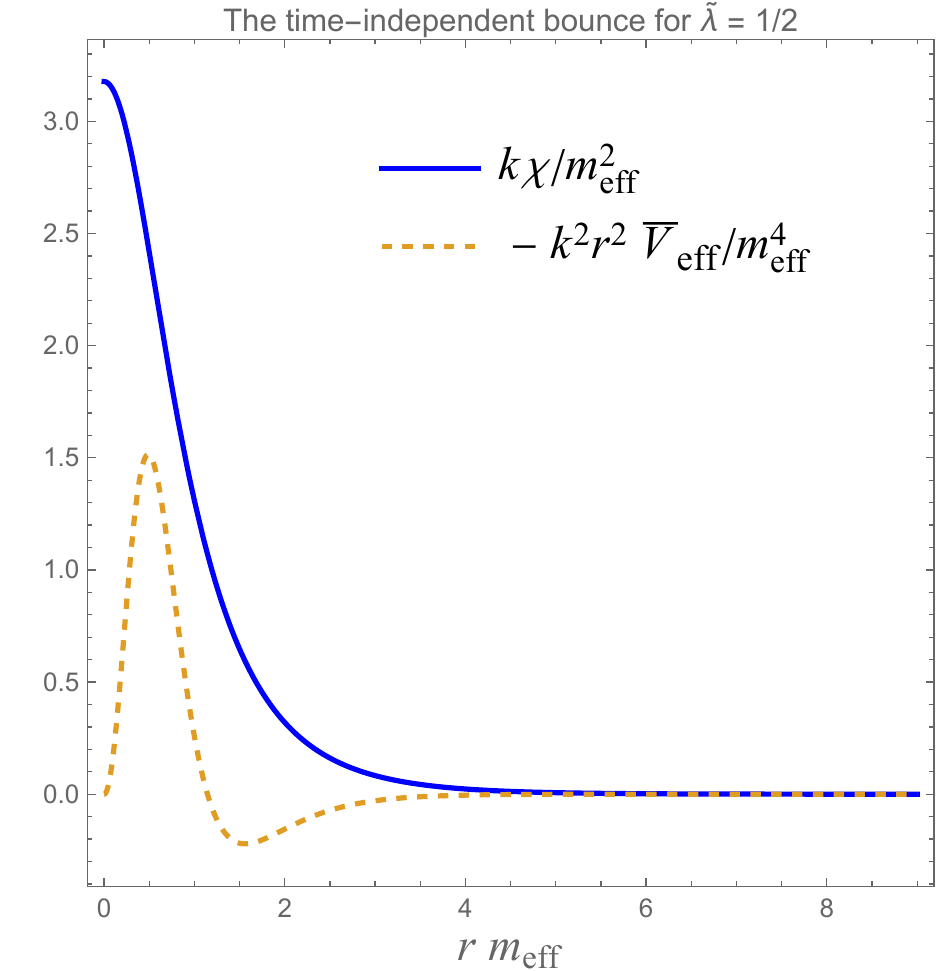}  \hspace{1cm} \includegraphics[scale=0.51]{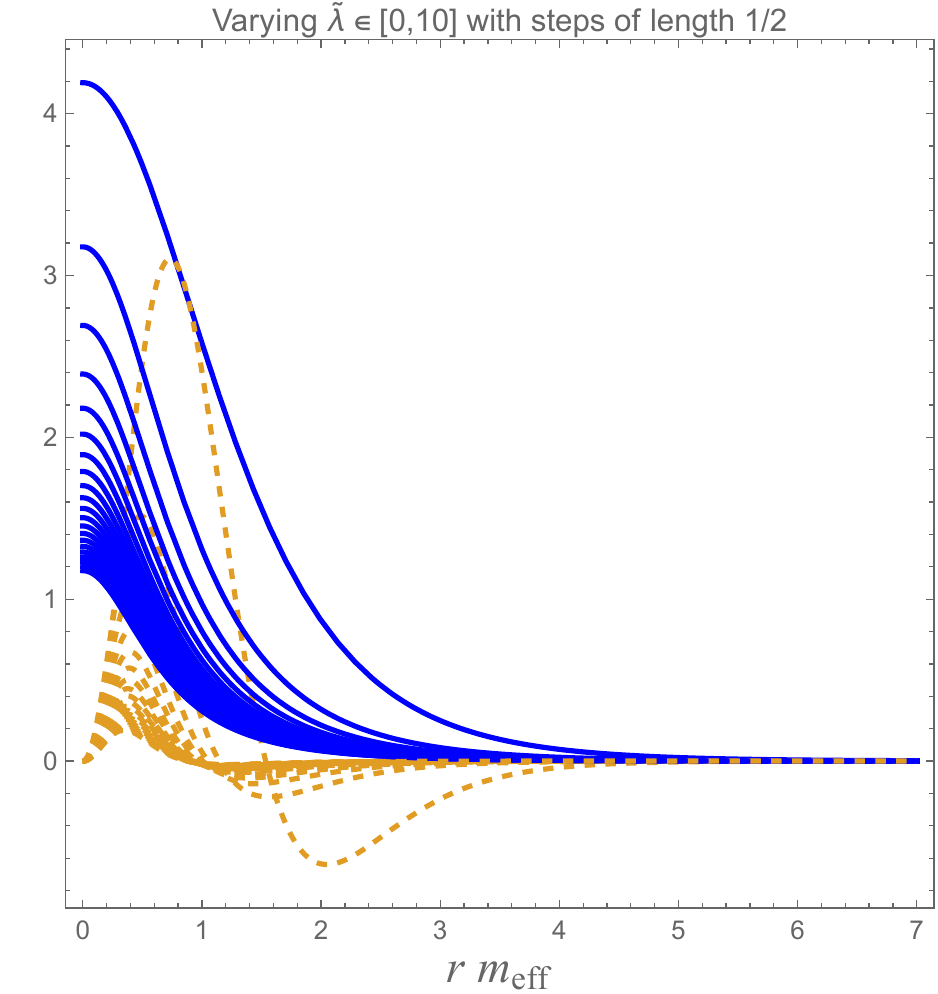}  
    \caption{\em  The relevant bounce and the corresponding integrand function (divided by $8\pi m_{\rm eff}^3/k^2$) appearing in the integral over $r m_{\rm eff}$ in~(\ref{bounceSd}) for the effective potential (\ref{barVnlo}) and varying the parameter $\tilde \lambda$ defined in~\eqref{lambdatDef}. The peak of the curves increases by decreasing $\tilde\lambda$.}\label{kappasetting}
  \end{center}
\end{figure}

 \begin{figure}[t!]
\includegraphics[scale=1.0]%[width=\linewidth,height=0.72\textheight,keepaspectratio]
{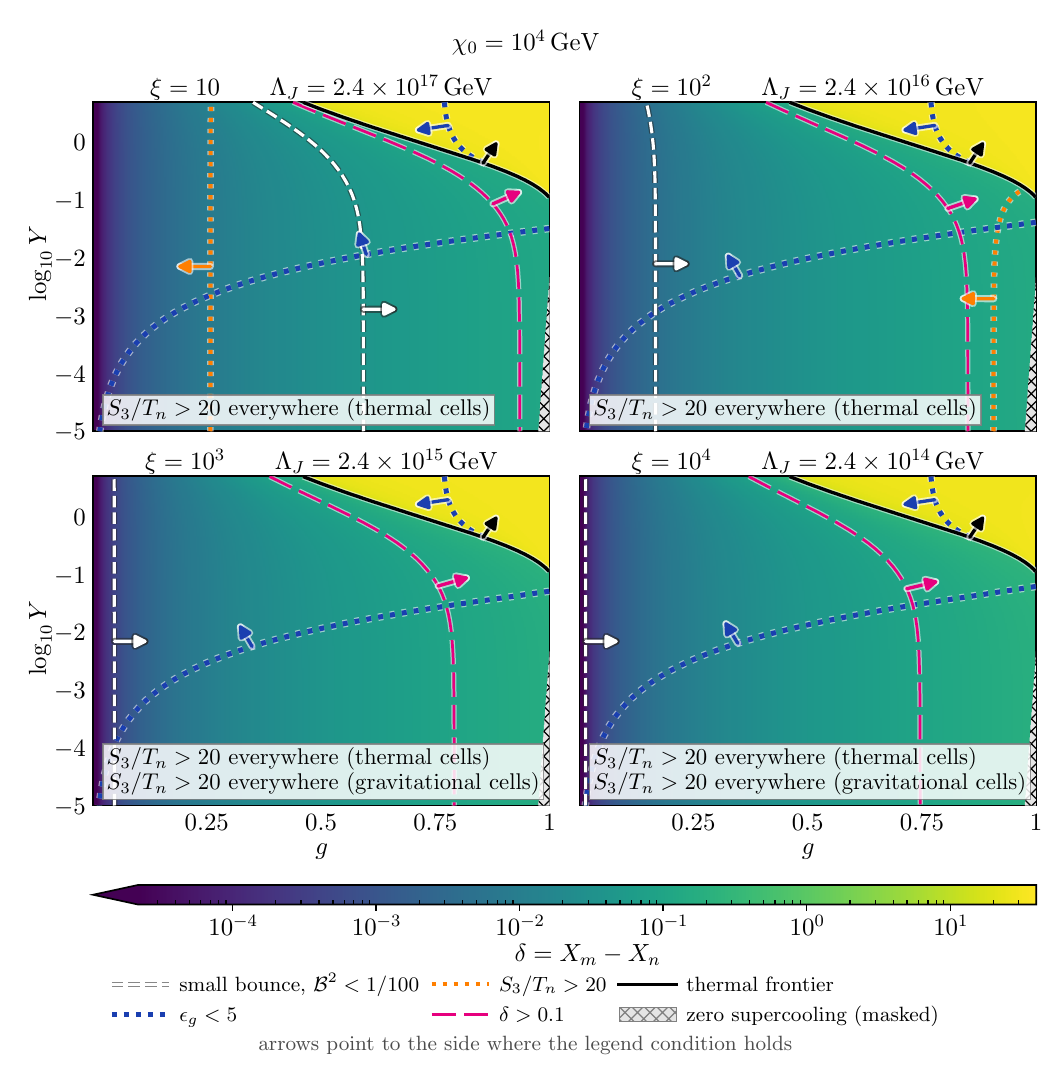}
\caption{\it Like Fig.~\ref{fig:LO_maps} but  in the improved supercool expansion at LO.  The lines with $\epsilon_g$ and $S_3/T_n$ constant are drawn separately on the thermal and on the gravitational cells, considering the solution for $T_n$ of each cell.}
\label{fig:imp_maps}
\end{figure}

Let us now consider the improved supercool expansion at LO (when $\bar V_{\rm eff}$ is approximated as in~\eqref{barVnlo}). We find that the fit 
\be S_3=a T\,\frac{\sqrt{1-s_\xi e^{2(X-X_m)}}}{X}\,Q\bigl(\tilde\lambda\bigr), \qquad Q(\tilde\lambda)\equiv\frac{27\pi}{2c_3}\,\frac{\tilde\lambda+\tilde\lambda e^{-1/\sqrt{\tilde\lambda}}}{1+\frac92\tilde\lambda}, \label{S3fit}
\ee
where 
\be \tilde\lambda(X)\equiv \frac{\lambda m^2_{\rm eff}}{k^2}=\frac{Yg^6}{12\tilde g^6}\,X\bigl(1-s_\xi e^{2(X-X_m)}\bigr), \label{lambdatDef}\ee 
reproduces at the $\sim1\%$ level $S_3$ determined by numerically computing the time-independent bounce varying $\tilde\lambda$
from very small up to very large values. Fig.~\ref{kappasetting} shows bounces obtained numerically varying $\tilde\lambda\in[0,10]$;  that figure explicitly shows that $R_b\sim1/m_{\rm eff}$; we also checked that the same is true for $\tilde\lambda> 10$. Using now the fit  in~\eqref{S3fit},  Eq.~\eqref{TnEq0} for $T_n$, which is valid also in the improved supercool expansion, reads (multiplying by $X$)
\be 4X^2-cX+a\sqrt{1- s_\xi e^{2(X-X_m)}} Q(\tilde\lambda(X))=0. \label{TnEqsbI}\ee 
Tables containing the numerical determination of $X_n$ from this equation can be downloaded from~\cite{dataset}.
  Like in the supercool expansion that we have previously discussed, here we are interested in the solution of Eq.~\eqref{TnEqsbI} with the smallest $X$, except $X=0$ because we obtained~\eqref{TnEqsbI} by multiplying Eq.~\eqref{TnEq0}  by $X$, and $X=0$ does not solve Eq.~\eqref{TnEq0}. Using~\eqref{beetaoH}, the inverse duration is then given by
\begin{equation}
\frac{\beta}{H_n}=\frac{a}{X_n}\left[\left(\frac{s_\xi e^{-2\delta}}{\sqrt{1-s_\xi e^{-2\delta}}}+\frac{\sqrt{1-s_\xi e^{-2\delta}}}{X_n}\right)Q(\tilde\lambda)-\sqrt{1-s_\xi e^{-2\delta}}Q'(\tilde\lambda(X_n))\,\tilde\lambda'(X_n)\right]-4,
\label{eq:betaH_imp}
\end{equation}
where $Q'$ and $\tilde\lambda'$ are the derivatives of $Q$ and $\tilde\lambda$ with respect to $\tilde\lambda$ and $X$ respectively. %Like in the supercool expansion, 
%here we cannot find an analytical  solution for small $\delta$ because $Q$ is not analytic at $\tilde\lambda=0$. However, 
%for sufficiently small $\delta$  and $\xi>0$ we still have a large $\beta/H_n$  because then $\epsilon_g$ is small (see Eq.~\eqref{epsilong2}) and therefore the (non-improved) supercool expansion is valid. 

\begin{figure}[t!]
\includegraphics[scale=.9]{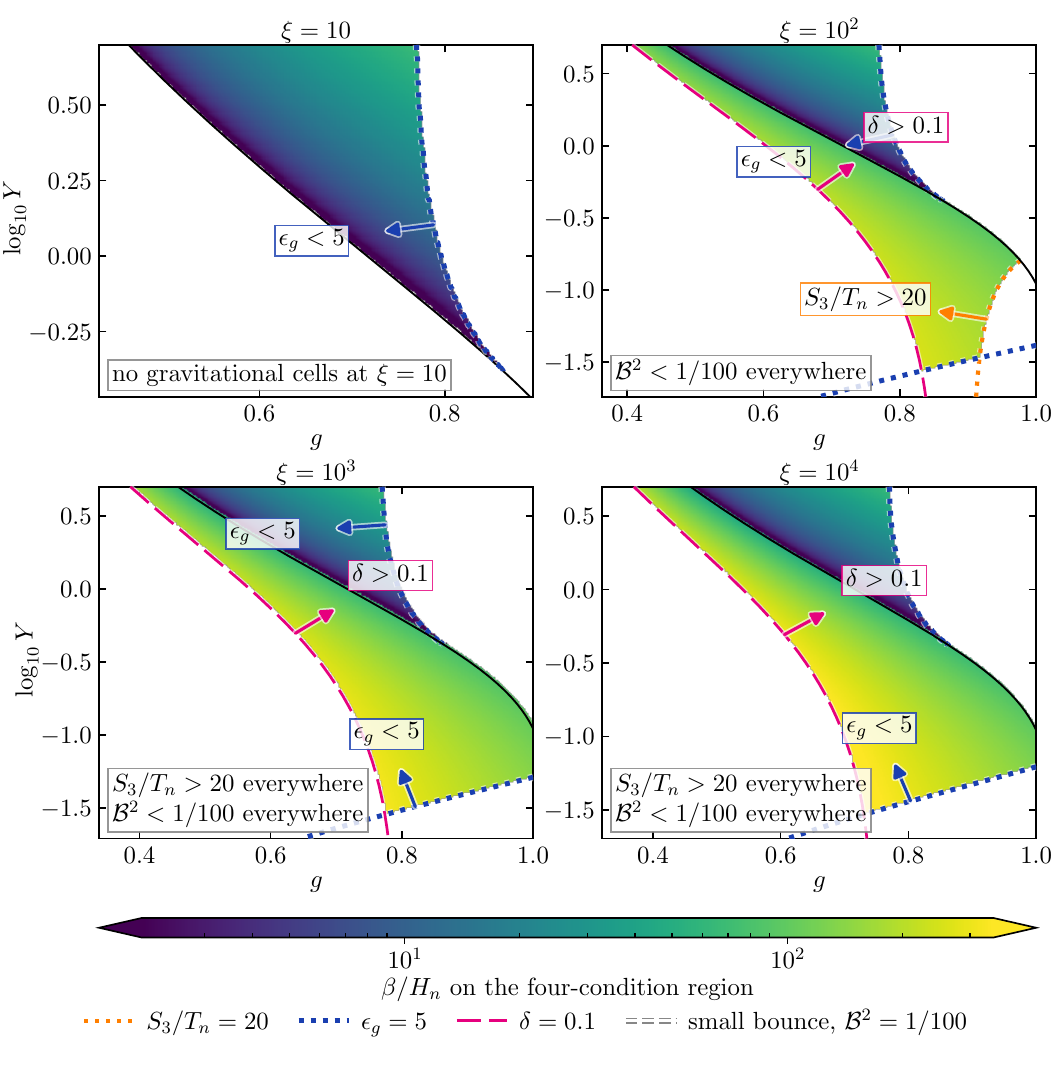}
\caption{\it $\beta/H_n$  in the improved supercool expansion at LO for $\chi_0=10^4\,$GeV. Gravitational cells lie to the left of the thermal frontier (thin black line), thermal cells to its right; both share the colour scale. The following further lines are shown: the boundaries $S_3/T_n=20$, $\epsilon_g=5$ and $\delta=0.1$ of the gravitational part, and the contour $\epsilon_g=5$ across the thermal cells, with arrows towards the side where they hold; the small-bounce condition $\mathcal B^2<1/100$ does not lead to further restrictions on the parameter space. At $\xi=10$ the thermal cells are shown, with the thermal frontier and the contour $\epsilon_g=5$.}
\label{fig:betaH_imp}
\end{figure}

Fig.~\ref{fig:imp_maps} gives the same information as Fig.~\ref{fig:LO_maps} but  in the improved supercool expansion at LO with $\tilde g = g$. We observe that our approximations hold in a larger region of the parameter space in the improved supercool expansion than in the (non-improved) supercool expansion. In Fig.~\ref{fig:betaH_imp} we then show the corresponding inverse duration of the PT. Like in the non-improved supercool expansion,  
we observe that a small $\delta$ enhances $\beta/H_n$  for $\xi>0$.

\begin{figure}[t!]
\includegraphics[scale=.9]{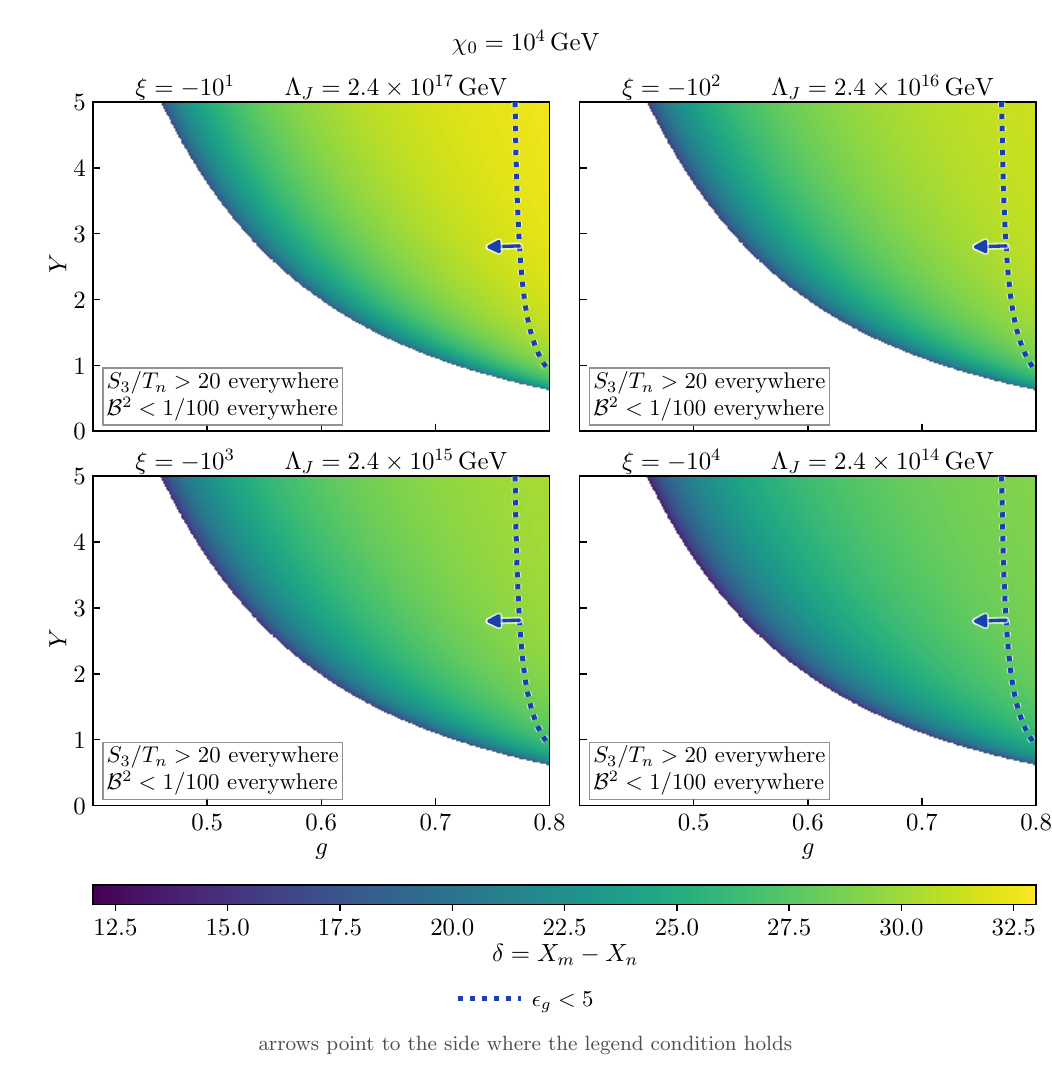}
\caption{\it 
 Nucleation temperature expressed in terms of $\delta$  in the improved supercool expansion at LO with gravity for $\xi<0$. All cells are thermal (there are no gravitational cells).}
\label{fig:negxi_delta}
\end{figure}

Fig.~\ref{fig:negxi_delta} shows the nucleation temperature in the improved supercool expansion at LO with gravity in the same setup, but  for $\xi<0$. Like in the (non-improved) supercool expansion, for $\xi<0$ gravity does not lead to further solutions for the nucleation temperature and this agrees with the fact that $\xi<0$ enhances the barrier in $\bar V_{eff}$.  We also find that where there is a solution the effect of $\xi$ is negligible (all cells are thermal) even taking $|\xi|$ up to $10^4$. We checked that this remains true also for much higher values of $\chi_0$ up to $10^{14}\,$GeV.

\subsection{Sizable Hubble rate}\label{Sizable Hubble rate}
Let us now turn to the case where the Hubble rate $H_I$ is sizable compared to $\pi m_{\rm eff}$:
\be H_I\gtrsim \pi m_{\rm eff}. \label{SizH}\ee
 We consider separately the three cases in which $\xi$ is negligibly small, $\xi$ is sizable and positive and  $\xi$ is sizable and negative.

Let us start from the case $\xi\approx0$ and assume first that the supercool expansion at LO  is applicable ($\bar V_{\rm eff}(\chi,T)$ can be approximated as in~\eqref{barVapp}). Since $\xi\approx0$, the condition in~\eqref{SizH} and the definition in~\eqref{meffDef} lead to
\be H_I\gtrsim \frac{g\pi T}{\sqrt{12}}. \label{SizH2}\ee 
On the other hand, Eq.~\eqref{HcrSE} tells us that in this case 
\be H_{\rm cr} =\frac{m}{\sqrt2} = \frac{g\, T}{\sqrt{24}}.\ee
Therefore, the condition in~\eqref{SizH2} tells us
\be H_I\gtrsim \sqrt2\pi H_{\rm cr}>H_{\rm cr}. \label{disH}\ee 
 When the supercool expansion at LO is not applicable and we need to use the improved one (to approximate $V_{\rm eff}$ as  in~\eqref{Vnlo}), the condition in~\eqref{SizH2} still holds, but for $\xi\approx0$  
%\be  H_{\rm cr}^2 \approx \frac{m^2}{2}+\frac{k^2}{8\lambda} -\frac{k}{8\lambda}\sqrt{k^2+4m^2\lambda }. \label{HcrISE2}\ee
the critical rate $H_{\rm cr}$ in the improved supercool expansion is always smaller than $H_{\rm cr}$ in the non-improved version, as we have seen around Eq.~\eqref{Hcrxi0}.   Therefore~\eqref{disH} remains valid. As a result, for $\xi\approx0$ the relevant solution (whose Euclidean action determines the false-vacuum decay rate) is the HM one of Sec.~\ref{Hawking-Moss solution}.

A similar conclusion also holds for $\xi$ sizable and positive as we now show. Recalling~\eqref{meffmT}, we note that the  characteristic bounce size in this case is significantly larger than $1/T$, which implies that the relevant solutions to determine  the false-vacuum decay rate are time independent. We now prove that these solutions should also be constant in space. Consider  the metric\footnote{That metric can be obtained from  de Sitter metric with Lorentzian signature~\cite{BirrellDavies}
\begin{equation}
ds_{\rm L}^2 = dt^2 - \frac{\cosh^2(tH)}{H^2}\, d\Omega^2
\end{equation}
%%%%For this expression of the de Sitter metric see Eq.~(5.65) of the Birrell-Davies  book (I find $\alpha=1/H$ with the RG \& TC - Code) and substitute $t\to -i\tau$ and then $\tau\to \tau-\pi /(2H)$.
by substituting the real time $t$ with $-i\tau$. Note that in the $H\to 0$ limit this Lorentzian metric tends to the flat spacetime as $\cosh^2(tH)\to 1$ and $d\Omega_3^2/H^2$ tends to the metric of the three-dimensional sphere with infinite radius.
} of the Euclidean de Sitter space,
\begin{equation}
ds^2 = d\tau^2 + \frac{\cos^2(H\tau)}{H^2}\, d\Omega^2, \label{dSmetric}
\end{equation}
where $H$ is the generic de Sitter Hubble rate, $\tau$ is the imaginary time of Euclidean spacetime  and $d\Omega^2$ is the metric of the unit three-dimensional sphere $S^3$. 
The part of the effective Euclidean action for $\chi$ which is relevant for our purpose is 
\begin{equation}
S_\chi = \int d^4 x \, \sqrt{|\det \hat g|}
\left[
\frac{1}{2} g^{\mu\nu} \partial_\mu \chi \, \partial_\nu \chi +\bar V_{\rm eff}(\chi, T)
\right],
\end{equation}
and the corresponding field equation is
\begin{equation}
\partial_\tau^2 \chi
- 3H\tan(H\tau)\, \partial_\tau \chi
+ \frac{H^2}{\cos^2(H\tau)} \nabla^2_{S^3} \chi
=\bar V_{\rm eff}^{(1)},
\end{equation}
where $\nabla^2_{S^{3}}$ is the Laplacian on $S^{3}$ and  $\bar V_{\rm eff}^{(1)}$ is the first derivative of $\bar V_{\rm eff}$ with respect to its first argument, $\chi$. For a time-independent solution, $\partial_\tau\chi=\partial_\tau^2\chi=0$, the equation above reduces to 
\begin{equation}
\nabla^2_{S^{3}} \chi
=\frac{\cos^2(H\tau)}{H^2} \bar V_{\rm eff}^{(1)}.
\end{equation}
Since $\chi$ is time independent this equation implies  that $\chi$ must be at the stationary point of $\bar V_{\rm eff}$ throughout space. The relevant solution is therefore the HM one that satisfies Eq.~\eqref{chiEqHMp}, which is nothing but $\bar V_{\rm eff}^{(1)}=0$.

Let us now turn to the case $\xi$ negative and  sizable, namely $|\xi|\gtrsim 1$. Note that in this situation 
\be m_{\rm eff}^2  = m^2 +12|\xi| H_I^2 \gg \frac{H_I^2}{\pi^2} \label{meffxin}\ee 
and therefore we recover the small-bounce regime of Sec.~\ref{Small Hubble rate regime}.
% Note that for $\xi<0$ \eqref{HcrISE} implies that $H_{\rm cr}$ is always bigger in the supercool expansion at LO than in the improved one. Therefore, the estimate in~\eqref{disH} continues to hold for $6\xi\ll 1$. When $6\xi \gtrsim 1$, we also have \xi\gg 1/(12 \pi^2)   and the conclusion in~\eqref{meffxin} holds.

We conclude that, when~\eqref{SizH} holds,   the relevant solution, which allows us to estimate the false-vacuum decay rate, is the HM one. 
%In this regime bubbles are  form. 
We will study the duration of the phase transition induced by the HM solution in Sec.~\ref{Number of e-folds}. Finally note that the regime in~\eqref{SizH} eventually occurs as the universe expands (and therefore $T$ decreases) when  there is no solution for $T_n$ and $\xi$ is not negative and  sizable.

\section{Gravitational-wave signals and detectability}\label{Gravitational-wave signals and detectability}

Let us consider now the GWs produced by the FOPTs described above. Here we study  GWs from collisions of vacuum bubbles, produced with the mechanisms of Sec.~\ref{Solutions with a  non-homogeneous scalar field}. This phenomenon is the typical source of GWs in our scenario. However, if  $\xi$ is so large and positive that supercooling is significantly reduced (because the barrier in $\bar V_{\rm eff}$ is significantly lowered) other sources of GWs, such as  turbulence and sound waves in the cosmic fluid, can be relevant. We will come back to this point below.

In the presence of supercooling (which leads to $\alpha\gg 1$) one finds the following GW spectrum 
due to vacuum bubble collisions\footnote{The spectral density $\Omega_{\rm GW}$ is defined  in the usual way
\be \Omega_{\rm GW}(f) \equiv \frac{f}{\rho_{\rm cr}}\frac{d\rho_{\rm GW}}{df},\ee 
where $\rho_{\rm cr}\equiv	3H_0^2\bp^2$  is the critical energy density, $H_0$ is the present value of the Hubble rate and $\rho_{\rm GW}$ is the energy density of the stochastic background.}~\cite{Caprini:2015zlo,Salvio:2023ext,Salvio:2023qgb}
\be \label{eq:gw_col} h^2 \Omega_{\rm GW}(f) \approx 1.29 
\tt 10^{-6}\left(\frac{H_r}{\beta}\right)^2\left(\frac{100}{g_*(T_r)}\right)^{1/3}\frac{3.8(f/f_{\rm peak})^{2.8}}{1+2.8(f/f_{\rm peak})^{3.8}}.\ee 
Here $T_r$ is the reheating temperature after supercooling, $H_r$ is the corresponding Hubble rate and $f_{\rm peak}$ is the red-shifted frequency peak today,  given by~\cite{Caprini:2015zlo} 
\be f_{\rm peak} \approx 3.79\, \frac{\beta}{H_r}\left( \frac{g_*(T_r)}{100}\right)^{1/6}\frac{T_r}{10^{8}{\rm GeV}}  \, {\rm Hz} . \ee

Reheating after supercooling in the RSB scenario was studied in~\cite{Rescigno:2025ong}, which found a large region of parameter space where reheating is fast, i.e.
\be T_r^4 \approx \frac{15 \bar\beta \chi_0^4}{8\pi^2 g_*(T_r)} \equiv T_{\rm eq}^4, \qquad \mbox{and}\qquad H_r\approx H_n \approx H_I,\label{TRHmax}\ee 
where $g_*(T)$ is the effective number of relativistic species at temperature $T$.  We will therefore assume fast reheating\footnote{If reheating is not fast $H_r$ and $T_r$ can depend on the details of the  specific model.}.  

 From~\eqref{eq:gw_col} it is clear that a large $\beta/H_n$ suppresses $\Omega_{\rm GW}$ and the same  is true for other sources of GWs (see e.g.~\cite{Caprini:2015zlo}). Therefore, a value of $\xi$ so large and positive (a value of $\delta$ so small) that the barrier in $\bar V_{\rm eff}$ is significantly lowered,
 % compared to the $\xi=0$ case
  such that other sources of GWs become relevant, corresponds to suppressed GW spectra because, as we have seen in Sec.~\ref{Small Hubble rate regime}, a small $\delta$ leads to a large $\beta/H_n$. Therefore, vacuum bubble collisions remain the main source of GWs in our scenario when we have the largest  $\Omega_{\rm GW}$ and the best chance of detectability.

  Note that a GW background acts as an extra radiation component and is highly constrained by big-bang nucleosynthesis (BBN) measurements of primordial elements~\cite{Boyle:2007zx,Stewart:2007fu,Kohri:2018awv}. This leads to a constraint on $\Omega_{\rm GW}$, which we call the BBN bound, that has been discussed in Ref.~\cite{Salvio:2023qgb}.

\begin{figure}[t!]
\includegraphics[scale=0.95]{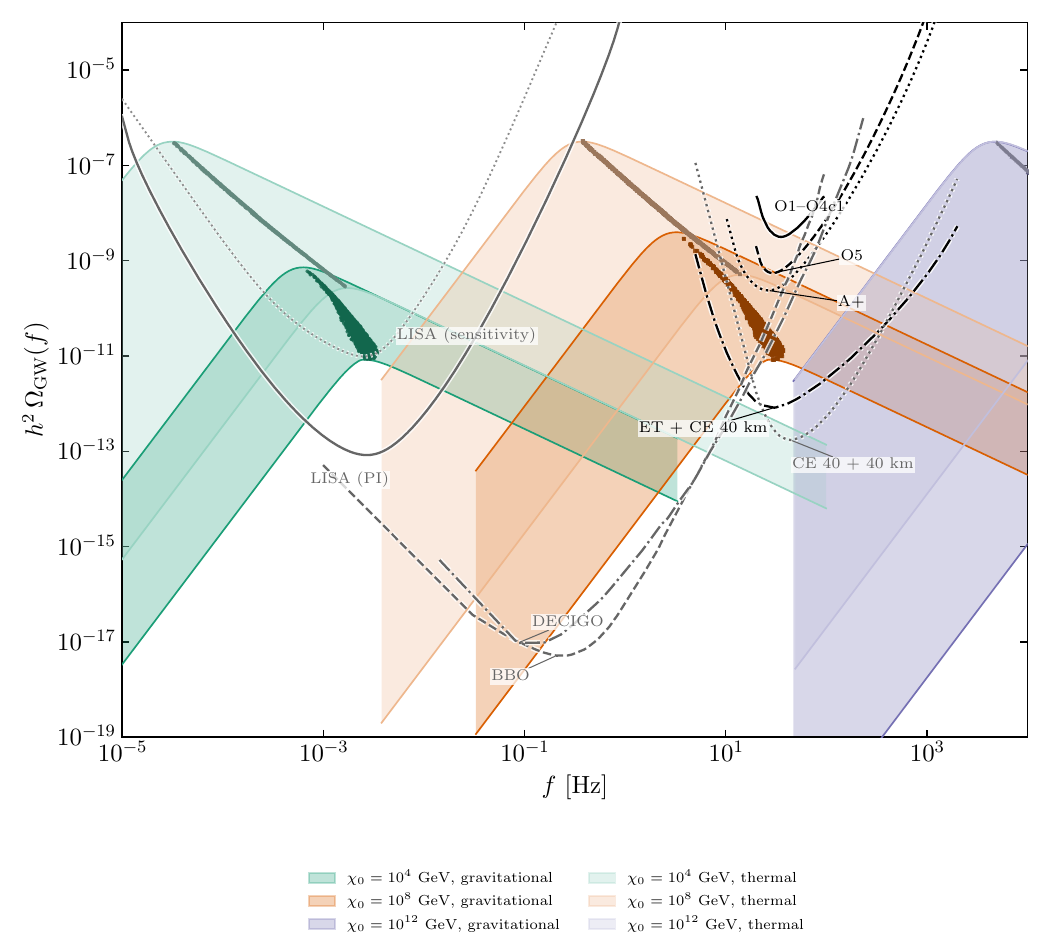}
\caption{\it   Theoretical predictions  for $\Omega_{\rm GW}$ obtained with  the improved supercool expansion at LO and compared with the sensitivities of several current and proposed GW detectors (see the text). The gravitational cells are in darker colors. The dots are the peaks of the predicted signals. Here we set 
$g_*(T_{\rm eq})= 110$. The other parameters are varied by keeping the four conditions in Fig.~\ref{fig:betaH_imp} and the BBN bound satisfied. We checked that the cutoff $\Lambda_J$ is much bigger than all relevant values of $\chi$ everywhere.}
\label{GWs}
\end{figure}

In Fig.~\ref{GWs} we show our theoretical predictions for $\Omega_{\rm GW}$ compared with the sensitivities of several current and proposed GW detectors: LIGO-Virgo-KAGRA runs O1-O4c1~\cite{LIGOScientific:2026mjf} and O5~\cite{LIGOO5}, LIGO A+~\cite{KAGRA:2021kbb,LIGOScientific:2023vdi}, LISA  sensitivity and power-law sensitivity (PI)~\cite{Babak:2021mhe}, an Einstein Telescope (ET) array combined with a Cosmic Explorer (CE) in the US (ET+CE 40\,km) and two CE detectors in the US with arm lengths  of 40 km (CE 40+40\,km)~\cite{Evans:2021gyd}, BBO and DECIGO (for power law spectra)~\cite{Thrane:2013oya,Dev:2019njv}. We observe that, although the thermal cells typically lead to larger values of $\Omega_{\rm GW}$, the gravitational cells are also within the reach of future GW detectors. In particular they can be probed by the LIGO-Virgo-KAGRA observations in the near future.

\begin{figure}[t!]\includegraphics[scale=0.9]{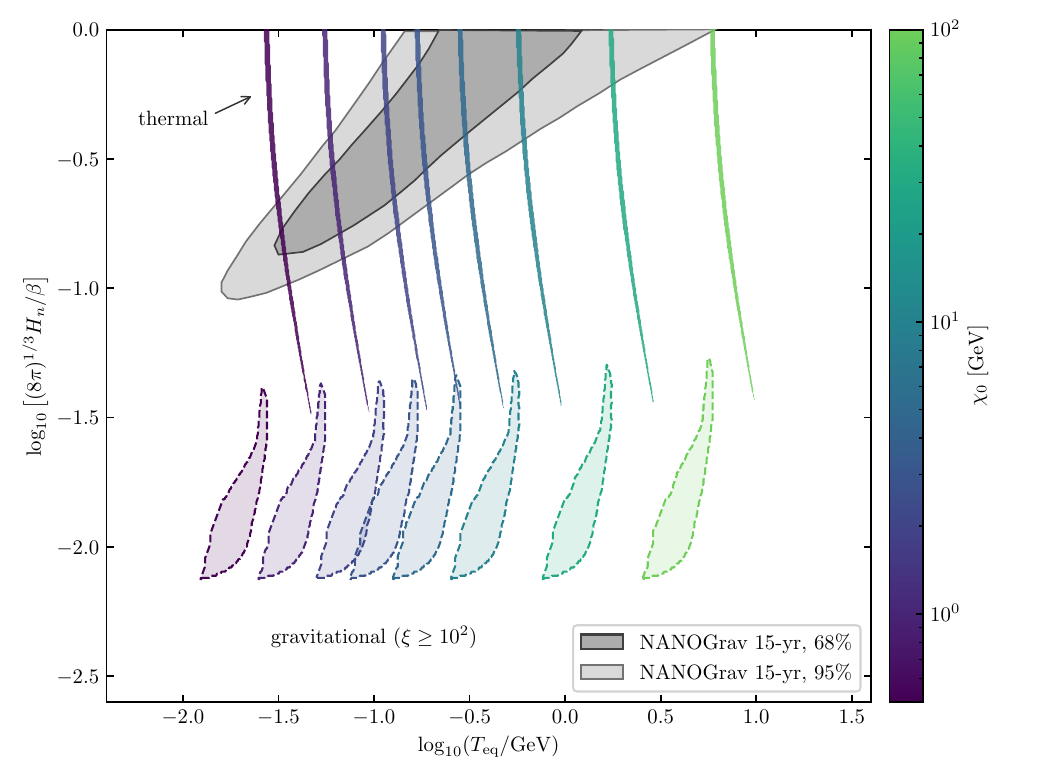}
\caption{\it 
Comparison between the theoretical predictions  for $\Omega_{\rm GW}$ obtained with  the improved supercool expansion at LO and the signals  detected by pulsar timing arrays. Here we set 
$g_*(T_{\rm eq})= 20$. The other parameters are varied by keeping the four conditions in Fig.~\ref{fig:betaH_imp} and the BBN bound satisfied. We checked that the cutoff $\Lambda_J$ is much bigger than all relevant values of $\chi$ everywhere.  The strips with darker colors  are the thermal
cells, while the regions with lighter colors and dashed boundaries are the gravitational cells (with $\xi \leq10^4$). 
 }
\label{fig:nanograv}
\end{figure}

In Fig.~\ref{fig:nanograv} we show our theoretical predictions for $\Omega_{\rm GW}$ compared with the signals  detected by pulsar timing arrays~\cite{NANOGrav:2023gor,NANOGrav:2023hvm}. We found that the signal is reproduced by the thermal cells independently of the values of positive $\xi\leq10^4$. The new parameter space allowed by such $\xi$ does not correspond to the detected 95\% regions.  

\section{Number of e-folds}\label{Number of e-folds}

   %checked english up to here
   
How long does thermal inflation last? This is an interesting question for the reason explained in the introduction, Sec.~\ref{Introduction}.
%as a prolonged phase of exponential expansion can have an important impact in cosmology. Indeed, the total number of e-folds of the universe's expansion is bounded from above (see e.g.~\cite{Liddle:2003as,WeinbergCosmo}).
 
When the relevant solutions (whose Euclidean actions determine the false-vacuum decay rate) are those with a  non-homogeneous scalar field introduced in Sec.~\ref{Solutions with a  non-homogeneous scalar field}, the number of e-folds $N_t$ of thermal inflation can be simply estimated as 
\be N_t\approx \log\left(\frac{T_{\text{eq}}}{T_n}\right), \ee
where, for a sufficiently large supercooling, we defined $T_{\text{eq}}$ in~\eqref{TRHmax} and $T_n$ can be determined with the model-independent approach of Sec.~\ref{Supercooled phase transitions}. Therefore, in this case, whenever there is a solution for $T_n$, 
\be N_t\approx \log\left(\frac{T_{\text{eq}}}{\chi_0}\right) +X_m-\delta, \ee
where we used the definitions in~\eqref{xDef}
%,~\eqref{XmDef} 
and~\eqref{deltaDef}. 
In the regions of the parameter space where there is no solution for $T_n$ the temperature keeps decreasing and eventually the HM solution becomes relevant, unless  $\xi$ is negative and sizable, as we have seen in Sec.~\ref{Sizable Hubble rate}. When $\xi$ is negative and sizable the PT does not complete if there is no solution for $T_n$. This remains true even when the temperature is so small that $\epsilon_g$ in~\eqref{epsilong} becomes large and the (improved) supercool expansion becomes inapplicable. Indeed, if so, the thermal contribution to $m_{\rm eff}$ is negligible compared with the gravitational one and $\Gamma$ would be at most
% when $|\xi|$ is very large the bounce size R_b\sim 1/m_eff = 1/\sqrt{12 |\xi|H^2} and therefore the pre-factor is of order [Buttazzo et al "Investigating ..."] (12|\xi|)^2H_I^4>> H_I^4, however when |\xi| increases the probability of false-vacuum decay must decrease because the barrier increases when \xi<0
 of order $H_I^4 e^{-S}\ll H_I^4$, where $S$ is the Euclidean action of the relevant zero-temperature bounce~\cite{Joti:2017fwe,Buttazzo:2013uya}. Therefore, in that case the bubbles are diluted by the expansion of the universe and they cannot collide.

When instead $\xi$ is not negative and the relevant solution is instead the HM of Sec.~\ref{Hawking-Moss solution}, which can occur when~\eqref{SizH} holds (see Sec.~\ref{Sizable Hubble rate}), 
 \be N_t\approx \log\left(\frac{T_{\text{eq}}}{T_*}\right)+N_{\rm HM}, \ee
where $T_*$ is the temperature at the cosmic time $t_*$ when~\eqref{SizH} starts to hold,
\be T_* \equiv \frac{\sqrt{12}\,H_I}{g}\sqrt{12\xi  + \frac{q }{\pi^2}}, \label{Tstar}\ee 
with $q$  a positive number of order $\approx 1$,
and $N_{\rm HM}$ is the number of e-folds  from $t_*$ up to the cosmic time when the probability $\wp$ of false-vacuum decay in a volume $V_I$ becomes sizable, $\wp\sim1$. 
Note that the HM solution only exists when there is  a barrier  in $\bar V_{\rm eff}$ therefore 
\be N_{\rm HM}< N_w, \label{NHMNw}\ee
where $N_w$ is the number of e-folds in the window of temperatures from $T_*$ (the value of the temperature at $t=t_*$)
down to the temperature 
\be T_{\rm canc} \equiv  \frac{12\sqrt{\xi}H_I}{g}\ee  at which the barrier  in $\bar V_{\rm eff}$ disappears
%\footnote{Here we do not consider the case in which $\xi$ is sizable and negative because, as we have seen in Sec.~\ref{Sizable Hubble rate}, that case is not compatible with~\eqref{SizH}. }:
\be N_w = \log\frac{T_*}{T_{\rm canc}} = \frac12\log\left(1+\frac{q}{12\pi^2 \xi}\right).\label{NwF}  \ee 
After the barrier has disappeared $\chi$ classically rolls towards the symmetry-breaking vacuum. The bound in~\eqref{NHMNw} and Eq.~\eqref{NwF} imply that $N_w$ and $N_{\rm HM}$  can be sizable only when $\xi$ is very small. 
%For example, for $N_w=2$ we obtain $\xi\lesssim qe^{-4}/(12\pi^2)$. 
We can therefore neglect $\xi$ in this situation.
% \sim e^{-22}
Here we start %the study of $N_{\rm HM}$ 
considering the case when $N_{\rm HM}$ is at least of order 10 because for $N_{\rm HM}\lesssim1$ we have a negligible impact on the cosmological predictions at the level of precision usually adopted in the literature on inflation.  During the period in which the $N_{\rm HM}$ e-folds elapse,  $T$ decreases by an exponentially large amount, $T=T_* \exp(-H_I (t-t_*))$
and the pre-factor (in the semi-classical expression for $\Gamma$) is of order $H_I^4$ when $\Gamma$ is sizable:
\be \Gamma \sim H_I^4\, e^{-S_{\rm HM}}. \label{GammaHM}\ee
There are situations where $T\gg H_I$ even for cosmic time $t>t_*$ and the pre-factor can be of order $T^4\gg H_I^4$, however, one then finds an exponential suppression of $\Gamma$ due to the fact that $S_{\rm HM}$ is then enhanced by the large ratio $T^4/H_I^4$. This can be explicitly checked in the (improved) supercool expansion at LO using Eqs.~\eqref{eq:SHM_geom},~\eqref{Vbctop} and~\eqref{chitop}.
%this corresponds to $g$ very small and/or $\xi$ is very large such that $T_*$ is enhanced  as $T_*$ is proportional to $1/g$ and grows as $\sqrt{\xi}$.

 The probability $d\wp$ of false-vacuum decay in a volume $V_I$ and during a cosmic-time interval $dt$ is 
 \be d\wp = \Gamma V_I dt. \ee 
 Taking $V_I = H_I^{-3}\exp(3N')$ and $dt = dN'/H_I$,  where $N'$ is a generic number of e-folds,  and assuming~\eqref{GammaHM}, the probability of false-vacuum decay in a volume $V_I$ after $N_{\rm HM}$ e-folds  is 
 \be \wp(N_{\rm HM})  \sim \int_0^{N_{\rm HM}} dN' e^{3N'} e^{-S_{\rm HM}}. \ee
 This integral is dominated by the $N'$ close to $N_{\rm HM}$ because $e^{3N'}$ exponentially grows and $S_{\rm HM}$ decreases\footnote{The fact that $S_{\rm HM}$ decreases can be shown in the (improved) supercool expansion at LO using again Eqs.~\eqref{eq:SHM_geom},~\eqref{Vbctop} and~\eqref{chitop} and the fact that $T$ decreases exponentially with $N'$.} as $N'$ grows. Therefore,
 \be \wp(N_{\rm HM}) \sim 
 %N_{\rm HM}
 e^{3N_{\rm HM}-S_{\rm HM}} \ee 
 and so to have a sizable probability we need
 \be N_{\rm HM} \approx \frac{S_{\rm HM}}{3}, \label{NHMeq}
 %\qquad (N_{\rm HM}\gtrsim 10), 
 \ee 
 where $S_{\rm HM}$ is evaluated at $N' = N_{\rm HM}$. Typically, this equation is not satisfied by $N_{\rm HM}$ of order 10 or larger. Then, from the time $t_*$ it takes generically  only a few e-folds for the HM solution to complete the transition. However, one should keep in mind that there are very special values of the parameters for which~\eqref{NHMeq} is satisfied by $N_{\rm HM}$ of order 10 or larger. Using %~\eqref{Tstar},
 Eqs.~\eqref{eq:SHM_geom},~\eqref{Vbctop} and~\eqref{chitop} and the supercool expansion at LO, for example, this happens when $T_*^4/(H_I^4Y) \gtrsim \exp(4N_{\rm HM})$, which leads to $T^4/(H_I^4Y)\gtrsim1$ after $N_{\rm HM}$ e-folds from $t_*$.

% , so when the relevant solution is the HM one
% \be N_t\approx \log\left(\frac{T_{\text{eq}}}{T_*}\right), \ee
% up to corrections of order no larger than one in $N_t$.

\section{Conclusions}\label{Conclusions}

In this work we have explored the role of gravity in the PTs associated with RSB through a model independent approach, which is valid when the quantity $\epsilon_g$ in Eq.~\eqref{epsilong} is small enough. This approach is based on the (improved) supercool expansion, which is valid for $\epsilon_g$ up to order one. 

Our analysis covers a general class of classically-scale-invariant models (with only dimensionless parameters at the classical level), including arbitrary numbers of scalars, fermions and vectors. These models can be interpreted as SM extensions or dark sectors weakly coupled to the SM. Crucially, we also included the non-minimal couplings parameterized by the  dimensionless quantities $\xi_{ab}$, which lead to a non-minimal coupling between gravity and the flat-direction field $\chi$, parameterized by the  dimensionless quantity $\xi$. This non-minimal coupling affects the effective mass $m_{\rm eff}$ of $\chi$ and therefore plays a crucial role: in our sign convention, a positive $\xi$ reduces the effective barrier felt by $\chi$, while a negative $\xi$ enhances the barrier. 

We relied on an effective-field-theory treatment of gravitational interactions which is valid only up to energy scales much below a cutoff. 

The Hawking-Moss configuration and the Coleman bounce are found to be relevant solutions (of the Euclidean field equations) in determining the false-vacuum decay rate per unit volume $\Gamma$ within the semi-classical approximation. We have studied these solutions when they give an important contribution to $\Gamma$, both in the supercool expansion at LO and in the improved version and including temperature effects. 

We have computed the nucleation temperature $T_n$ when the bounce solutions dominate $\Gamma$. In the supercool expansion, $T_n$ can be determined analytically taking into account $\xi$.  In the improved version at LO the bounce action is only known through a fit and we have therefore determined $T_n$ only numerically, when it exists,  varying the model-independent parameters within physically motivated and consistent ranges. The produced data can be downloaded at~\cite{dataset}.  We have also explained how to determine the duration of the FOPT when $T_n$ exists.  
A positive (negative) value of $\xi$ enlarges (reduces) the region of  parameter space where $T_n$ exists.

The enlarged parameter space for $\xi>0$ leads to new predictions for the GW spectra induced by the FOPT, which are genuine manifestations  of spacetime curvature effects in the FOPT. These predictions can be tested through future GW  detectors. In particular they will be probed soon by the LIGO-Virgo-KAGRA observations.
On the other hand, the spacetime curvature  plays typically a negligible role in the FOPT in the regions of parameter space where the  background of GWs detected by the PTAs can be reproduced. 
 
 Finally, we have estimated in a model-independent manner the number $N_t$ of e-folds during thermal inflation,  when $\Gamma$ is dominated by the Coleman bounce or by the Hawking-Moss solution. A sizable value of $N_t$ has an important impact on cosmology, which has to be taken into account in classically-scale-invariant models.

 \subsubsection*{Acknowledgments}
We thank Francesco Rescigno for useful discussions.

%\appendix

 \vspace{1cm}
 %\newpage
\footnotesize
\begin{multicols}{2}

\end{multicols}


\begin{thebibliography}{}
\small{

 \bibitem{Abbott:2016blz} 
  B.~P.~Abbott {\it \textit{et al.}} [LIGO Scientific and Virgo Collaborations],
  ``Observation of Gravitational Waves from a Binary Black Hole Merger,''
  Phys.\ Rev.\ Lett.\  {\bf 116}, 061102 (2016)
 % [Phys.\ Rev.\ Lett.\  {\bf 116}, 061102 (2016)]
  doi:10.1103/PhysRevLett.116.061102
  [\hhref{1602.03837}].
  
  
  %\cite{TheLIGOScientific:2016wyq1}
\bibitem{TheLIGOScientific:2016wyq1}
B.~Abbott \textit{\textit{et al.}} [LIGO Scientific and Virgo],
``GW150914: Implications for the stochastic gravitational wave background from binary black holes,''
Phys. Rev. Lett. \textbf{116}, no.13, 131102 (2016)
doi:10.1103/PhysRevLett.116.131102
[\hhref{1602.03847}].

%\cite{LIGOScientific:2017ync}
\bibitem{LIGOScientific:2017ync}
B.~P.~Abbott \textit{et al.}
``Multi-messenger Observations of a Binary Neutron Star Merger,''
Astrophys. J. Lett. \textbf{848} (2017) no.2, L12
doi:10.3847/2041-8213/aa91c9
[\hhref{1710.05833}].


 
%\cite{NANOGrav:2023gor}
\bibitem{NANOGrav:2023gor}
G.~Agazie \textit{et al.} [NANOGrav],
``The NANOGrav 15 yr Data Set: Evidence for a Gravitational-wave Background,''
Astrophys. J. Lett. \textbf{951} (2023) no.1, L8
doi:10.3847/2041-8213/acdac6
[\hhref{2306.16213}].

%\cite{NANOGrav:2023hvm}
\bibitem{NANOGrav:2023hvm}
A.~Afzal \textit{et al.} [NANOGrav],
``The NANOGrav 15 yr Data Set: Search for Signals from New Physics,''
Astrophys. J. Lett. \textbf{951}, no.1, L11 (2023)
[erratum: Astrophys. J. Lett. \textbf{971}, no.1, L27 (2024)]
doi:10.3847/2041-8213/acdc91
[\hhref{2306.16219}].

%\cite{Antoniadis:2023ott}
\bibitem{Antoniadis:2023ott}
J.~Antoniadis \textit{et al.} [EPTA and InPTA],
``The second data release from the European Pulsar Timing Array - III. Search for gravitational wave signals,''
Astron. Astrophys. \textbf{678}, A50 (2023)
doi:10.1051/0004-6361/202346844
[\hhref{2306.16214}].

%\cite{Reardon:2023gzh}
\bibitem{Reardon:2023gzh}
D.~J.~Reardon, A.~Zic, R.~M.~Shannon, G.~B.~Hobbs, M.~Bailes, V.~Di Marco, A.~Kapur, A.~F.~Rogers, E.~Thrane and J.~Askew, \textit{et al.}
``Search for an Isotropic Gravitational-wave Background with the Parkes Pulsar Timing Array,''
Astrophys. J. Lett. \textbf{951} (2023) no.1, L6
doi:10.3847/2041-8213/acdd02
[\hhref{2306.16215}].

%\cite{Xu:2023wog}
\bibitem{Xu:2023wog}
H.~Xu, S.~Chen, Y.~Guo, J.~Jiang, B.~Wang, J.~Xu, Z.~Xue, R.~N.~Caballero, J.~Yuan and Y.~Xu, \textit{et al.}
``Searching for the Nano-Hertz Stochastic Gravitational Wave Background with the Chinese Pulsar Timing Array Data Release I,''
Res. Astron. Astrophys. \textbf{23} (2023) no.7, 075024
doi:10.1088/1674-4527/acdfa5
[\hhref{2306.16216}].


%\cite{Hawking:1982ga}
\bibitem{Hawking:1982ga}
S.~W.~Hawking, I.~G.~Moss and J.~M.~Stewart,
``Bubble Collisions in the Very Early Universe,''
Phys. Rev. D \textbf{26} (1982), 2681
doi:10.1103/PhysRevD.26.2681

%\cite{Crawford:1982yz}
\bibitem{Crawford:1982yz}
M.~Crawford and D.~N.~Schramm,
``Spontaneous Generation of Density Perturbations in the Early Universe,''
Nature \textbf{298} (1982), 538-540
doi:10.1038/298538a0

%\cite{Kodama:1982sf}
\bibitem{Kodama:1982sf}
H.~Kodama, M.~Sasaki and K.~Sato,
``Abundance of Primordial Holes Produced by Cosmological First Order Phase Transition,''
Prog. Theor. Phys. \textbf{68} (1982), 1979
doi:10.1143/PTP.68.1979

%\cite{Moss:1994iq}
\bibitem{Moss:1994iq}
I.~G.~Moss,
``Singularity formation from colliding bubbles,''
Phys. Rev. D \textbf{50} (1994), 676-681
doi:10.1103/PhysRevD.50.676


%\cite{Khlopov:1999ys}
\bibitem{Khlopov:1999ys}
M.~Y.~Khlopov, R.~V.~Konoplich, S.~G.~Rubin and A.~S.~Sakharov,
``First order phase transitions as a source of black holes in the early universe,''
Grav. Cosmol. \textbf{2} (1999), S1
[\hhref{hep-ph/9912422}].


%\cite{Maggiore:2018sht}
\bibitem{Maggiore:2018sht}
M.~Maggiore,
``Gravitational Waves. Vol. 2: Astrophysics and Cosmology,'' Oxford University Press, 3, 2018.

%\cite{Kawana:2021tde}
\bibitem{Kawana:2021tde}
K.~Kawana and K.~P.~Xie,
``Primordial black holes from a cosmic phase transition: The collapse of Fermi-balls,''
Phys. Lett. B \textbf{824} (2022), 136791
doi:10.1016/j.physletb.2021.136791
[\hhref{2106.00111}].

%\cite{Liu:2021svg}
\bibitem{Liu:2021svg}
J.~Liu, L.~Bian, R.~G.~Cai, Z.~K.~Guo and S.~J.~Wang,
``Primordial black hole production during first-order phase transitions,''
Phys. Rev. D \textbf{105} (2022) no.2, L021303
doi:10.1103/PhysRevD.105.L021303
[\hhref{2106.05637}].

%\cite{Baker:2021sno}
\bibitem{Baker:2021sno}
M.~J.~Baker, M.~Breitbach, J.~Kopp and L.~Mittnacht,
``Detailed Calculation of Primordial Black Hole Formation During First-Order Cosmological Phase Transitions,'' Phys. Rev. D \textbf{111} (2025) no.6, 063544
doi:10.1103/PhysRevD.111.063544
[\hhref{2110.00005}].

%\cite{Kawana:2022olo}
\bibitem{Kawana:2022olo}
K.~Kawana, T.~Kim and P.~Lu,
``PBH Formation from Overdensities in Delayed Vacuum Transitions,'' Phys. Rev. D \textbf{108} (2023) no.10, 103531
doi:10.1103/PhysRevD.108.103531
[\hhref{2212.14037}].

%\cite{Athron:2023xlk}
\bibitem{Athron:2023xlk}
P.~Athron, C.~Bal{\'a}zs, A.~Fowlie, L.~Morris and L.~Wu,
``Cosmological phase transitions: From perturbative particle physics to gravitational waves,''
Prog. Part. Nucl. Phys. \textbf{135} (2024), 104094
doi:10.1016/j.ppnp.2023.104094
[\hhref{2305.02357}].

%\cite{Lewicki:2023ioy}
\bibitem{Lewicki:2023ioy}
M.~Lewicki, P.~Toczek and V.~Vaskonen,
``Primordial black holes from strong first-order phase transitions,''
JHEP \textbf{09} (2023), 092
doi:10.1007/JHEP09(2023)092
[\hhref{2305.04924}].

%\cite{Gouttenoire:2023naa}
\bibitem{Gouttenoire:2023naa}
Y.~Gouttenoire and T.~Volansky,
``Primordial black holes from supercooled phase transitions,''
Phys. Rev. D \textbf{110} (2024) no.4, 043514
doi:10.1103/PhysRevD.110.043514
[\hhref{2305.04942}].

%\cite{Ahriche:2023jdq}
\bibitem{Ahriche:2023jdq}
A.~Ahriche, S.~Kanemura and M.~Tanaka,
``Gravitational waves from phase transitions in scale invariant models,''
JHEP \textbf{01} (2024), 201
doi:10.1007/JHEP01(2024)201
[\hhref{2308.12676}].

%\cite{Hashino:2025fse}
\bibitem{Hashino:2025fse}
K.~Hashino, S.~Kanemura, T.~Takahashi, M.~Tanaka and C.~M.~Yoo,
``Super-critical primordial black hole formation via delayed first-order electroweak phase transition,''
[\hhref{2501.11040}].

%\cite{Kierkla:2025vwp}
\bibitem{Kierkla:2025vwp}
M.~Kierkla, N.~Ramberg, P.~Schicho and D.~Schmitt,
``Thermodynamical uncertainties for primordial black holes from cosmological phase transitions,''
Phys. Rev. D \textbf{113}, no.9, 9 (2026)
doi:10.1103/nj27-ltyg
[\hhref{2506.15496}].

%\cite{Witten:1980ez}
\bibitem{Witten:1980ez}
E.~Witten,
``Cosmological Consequences of a Light Higgs Boson,''
Nucl. Phys. B \textbf{177} (1981), 477-488
doi:10.1016/0550-3213(81)90182-6

 %\cite{Salvio:2023qgb}
\bibitem{Salvio:2023qgb}
A.~Salvio,
``Model-independent radiative symmetry breaking and gravitational waves,''
JCAP \textbf{04} (2023), 051
doi:10.1088/1475-7516/2023/04/051
[\hhref{2302.10212}].



%\cite{Salvio:2023ext}
	\bibitem{Salvio:2023ext}
	A.~Salvio,
	``Supercooling in radiative symmetry breaking: theory extensions, gravitational wave detection and primordial black holes,''
	JCAP \textbf{12} (2023), 046
	doi:10.1088/1475-7516/2023/12/046
	[\hhref{2307.04694}].
	
	

	%\cite{Banerjee:2024cwv}
\bibitem{Banerjee:2024cwv}
I.~K.~Banerjee, F.~Rescigno and A.~Salvio,
``Primordial black holes (as dark matter) from the supercooled phase transitions with radiative symmetry breaking,''
JCAP \textbf{07} (2025), 007
doi:10.1088/1475-7516/2025/07/007
[\hhref{2412.06889}].



%\cite{Salvio:2024upo}
\bibitem{Salvio:2024upo}
A.~Salvio,
``Introduction to Thermal Field Theory,'' Universe \textbf{11} (2025) no.1, 16
doi:10.3390/universe11010016
[\hhref{2411.02498}].

%\cite{Salvio:2026bco}
\bibitem{Salvio:2026bco}
A.~Salvio,
``Supercooled phase transitions with radiative symmetry breaking,''
Int. J. Mod. Phys. A \textbf{41} (2026) no.24n25, 2646004
doi:10.1142/s0217751x26460048
[\hhref{2602.20246}].

%\cite{Coleman:1973jx}
\bibitem{Coleman:1973jx}
S.~R.~Coleman and E.~J.~Weinberg,
``Radiative Corrections as the Origin of Spontaneous Symmetry Breaking,''
Phys. Rev. D \textbf{7} (1973), 1888-1910
doi:10.1103/PhysRevD.7.1888

%\cite{Gildener:1976ih}
\bibitem{Gildener:1976ih}
E.~Gildener and S.~Weinberg,
``Symmetry Breaking and Scalar Bosons,''
Phys. Rev. D \textbf{13} (1976), 3333
doi:10.1103/PhysRevD.13.3333


%\cite{Salvio:2023blb}
\bibitem{Salvio:2023blb}
A.~Salvio,
``Pulsar timing arrays and primordial black holes from a supercooled phase transition,''
Phys. Lett. B \textbf{852} (2024), 138639
doi:10.1016/j.physletb.2024.138639
[\hhref{2312.04628}].

%\cite{Cutrona:2026mxm}
\bibitem{Cutrona:2026mxm}
F.~Cutrona, F.~Rescigno and A.~Salvio,
``ACT-Planck data and phase transitions from a viable no-scale Standard Model completion,''
Phys. Rev. D \textbf{114} (2026) no.4, 043513
doi:10.1103/38mz-pw8t
[\hhref{2603.24664}].


%\cite{Kim:2026zna}
\bibitem{Kim:2026zna}
T.~Kim and W.~I.~Park,
``Primordial black holes as a natural consequence of scale-invariance,''
[\hhref{2610.10677}].

%\cite{Fujikura:2023lkn}
\bibitem{Fujikura:2023lkn}
K.~Fujikura, S.~Girmohanta, Y.~Nakai and M.~Suzuki,
``NANOGrav signal from a dark conformal phase transition,''
Phys. Lett. B \textbf{846}, 138203 (2023)
doi:10.1016/j.physletb.2023.138203
[\hhref{2306.17086}].

%\cite{Gouttenoire:2023bqy}
\bibitem{Gouttenoire:2023bqy}
Y.~Gouttenoire,
``First-Order Phase Transition Interpretation of Pulsar Timing Array Signal Is Consistent with Solar-Mass Black Holes,''
Phys. Rev. Lett. \textbf{131}, no.17, 17 (2023)
doi:10.1103/PhysRevLett.131.171404
[\hhref{2307.04239}].

%\cite{Puchades-Ibanez:2026ksh}
\bibitem{Puchades-Ibanez:2026ksh}
C.~Puchades-Ib{\'a}{\~n}ez and P.~Schwaller,
``The devil in the transition: NLO nucleation and the particle physics behind the PTA signal,''
[\hhref{2609.12098}].

%\cite{Joti:2017fwe}
\bibitem{Joti:2017fwe}
A.~Joti, A.~Katsis, D.~Loupas, A.~Salvio, A.~Strumia, N.~Tetradis and A.~Urbano,
``(Higgs) vacuum decay during inflation,''
JHEP \textbf{07} (2017), 058
doi:10.1007/JHEP07(2017)058
[\hhref{1706.00792}].

%\cite{Rajantie:2017ajw}
\bibitem{Rajantie:2017ajw}
A.~Rajantie and S.~Stopyra,
``Standard Model vacuum decay in a de Sitter Background,''
Phys. Rev. D \textbf{97}, no.2, 025012 (2018)
doi:10.1103/PhysRevD.97.025012
[\hhref{1707.09175}].

%\cite{Markkanen:2018pdo}
\bibitem{Markkanen:2018pdo}
T.~Markkanen, A.~Rajantie and S.~Stopyra,
``Cosmological Aspects of Higgs Vacuum Metastability,''
Front. Astron. Space Sci. \textbf{5}, 40 (2018)
doi:10.3389/fspas.2018.00040
[\hhref{1809.06923}].


%
\bibitem{Linde:1980tt}
A.~D.~Linde,
``Fate of the False Vacuum at Finite Temperature: Theory and Applications,''
Phys. Lett. B \textbf{100} (1981), 37-40 doi:10.1016/0370-2693(81)90281-1

%\cite{Linde:1981zj}
\bibitem{Linde:1981zj}
A.~D.~Linde,
``Decay of the False Vacuum at Finite Temperature,''
Nucl. Phys. B \textbf{216} (1983), 421. Erratum: [Nucl. Phys. B \textbf{223}, 544 (1983)] doi:10.1016/0550-3213(83)90072-X


%\cite{Coleman:1977py}
\bibitem{Coleman:1977py}
S.~R.~Coleman,
``The Fate of the False Vacuum. 1. Semiclassical Theory,''
Phys. Rev. D \textbf{15} (1977), 2929-2936
[erratum: Phys. Rev. D \textbf{16} (1977), 1248]
doi:10.1103/PhysRevD.15.2929


%\cite{Coleman:1980aw}
\bibitem{Coleman:1980aw}
S.~R.~Coleman and F.~De Luccia,
``Gravitational Effects on and of Vacuum Decay,''
Phys. Rev. D \textbf{21} (1980), 3305
doi:10.1103/PhysRevD.21.3305


%\cite{Hawking:1981fz}
\bibitem{Hawking:1981fz}
S.~W.~Hawking and I.~G.~Moss,
``Supercooled Phase Transitions in the Very Early Universe,''
Phys. Lett. B \textbf{110} (1982), 35-38
doi:10.1016/0370-2693(82)90946-7


%\cite{Liddle:2003as}
\bibitem{Liddle:2003as}
A.~R.~Liddle and S.~M.~Leach,
``How long before the end of inflation were observable perturbations produced?,''
Phys. Rev. D \textbf{68} (2003), 103503
[\hhref{astro-ph/0305263}].

\bibitem{WeinbergCosmo}
S.~Weinberg, ``Cosmology," Oxford University Press, Oxford, 2008.


\bibitem{BirrellDavies}
N.D. Birrell and P.C.W. Davies, ``Quantum Fields in
Curved Space" (Cambridge, UK: Univ. Pr., 1982).

%\cite{Salvio:2014soa}
\bibitem{Salvio:2014soa}
A.~Salvio and A.~Strumia,
``Agravity,''
JHEP \textbf{06} (2014), 080
doi:10.1007/JHEP06(2014)080
[\hhref{1403.4226}].

%\cite{Kannike:2015apa}
\bibitem{Kannike:2015apa}
K.~Kannike, G.~H\"utsi, L.~Pizza, A.~Racioppi, M.~Raidal, A.~Salvio and A.~Strumia,
``Dynamically Induced Planck Scale and Inflation,''
JHEP \textbf{05} (2015), 065
doi:10.1007/JHEP05(2015)065
[\hhref{1502.01334}].

%\cite{Salvio:2017qkx}
\bibitem{Salvio:2017qkx}
A.~Salvio and A.~Strumia,
``Agravity up to infinite energy,''
Eur. Phys. J. C \textbf{78} (2018) no.2, 124
doi:10.1140/epjc/s10052-018-5588-4
[\hhref{1705.03896}].

%\cite{Salvio:2017xul}
\bibitem{Salvio:2017xul}
A.~Salvio,
``Inflationary Perturbations in No-Scale Theories,''
Eur. Phys. J. C \textbf{77} (2017) no.4, 267
doi:10.1140/epjc/s10052-017-4825-6
[\hhref{1703.08012}].

%\cite{Salvio:2019wcp}
\bibitem{Salvio:2019wcp}
A.~Salvio,
``Quasi-Conformal Models and the Early Universe,''
Eur. Phys. J. C \textbf{79} (2019) no.9, 750
doi:10.1140/epjc/s10052-019-7267-5
[\hhref{1907.00983}].

%\cite{Salvio:2024joi}
\bibitem{Salvio:2024joi}
A.~Salvio,
``A non-perturbative and background-independent formulation of quadratic gravity,''
JCAP \textbf{07}, 092 (2024)
doi:10.1088/1475-7516/2024/07/092
[\hhref{2404.08034}].


%\cite{Salvio:2020axm}
\bibitem{Salvio:2020axm}
A.~Salvio,
``Dimensional Transmutation in Gravity and Cosmology,''
Int. J. Mod. Phys. A \textbf{36} (2021) no.08n09, 2130006
doi:10.1142/S0217751X21300064
[\hhref{2012.11608}].


\bibitem{BurgessLeeTrott} C.~P.~Burgess, H.~M.~Lee and M.~Trott, ``Power-counting and the Validity of the Classical Approximation During Inflation,'' JHEP {\bf 09} (2009) 103 [\hhref{0902.4465}].

\bibitem{BarbonEspinosa} J.~L.~F.~Barbon and J.~R.~Espinosa, ``On the Naturalness of Higgs Inflation,'' Phys.\ Rev.\ D {\bf 79} (2009) 081302 [\hhref{0903.0355}].

%\cite{Burgess:2010zq}
\bibitem{Burgess:2010zq}
C.~P.~Burgess, H.~M.~Lee and M.~Trott,
``Comment on Higgs Inflation and Naturalness,''
JHEP \textbf{07}, 007 (2010)
doi:10.1007/JHEP07(2010)007
[\hhref{1002.2730}].

\bibitem{Hertzberg} M.~P.~Hertzberg, ``On Inflation with Non-minimal Coupling,'' JHEP {\bf 11} (2010) 023 [\hhref{1002.2995}].

%\cite{Burgess:2014lza}
\bibitem{Burgess:2014lza}
C.~P.~Burgess, S.~P.~Patil and M.~Trott,
``On the Predictiveness of Single-Field Inflationary Models,''
JHEP \textbf{06}, 010 (2014)
doi:10.1007/JHEP06(2014)010
[\hhref{1402.1476}].

\bibitem{BezrukovMagnin} F.~Bezrukov, A.~Magnin, M.~Shaposhnikov and S.~Sibiryakov, ``Higgs inflation: consistency and generalisations,'' JHEP {\bf 01} (2011) 016 [\hhref{1008.5157}].



%\cite{Rescigno:2025ong}
\bibitem{Rescigno:2025ong}
F.~Rescigno and A.~Salvio,
``Reheating after the supercooled phase transitions with radiative symmetry breaking,''
JCAP \textbf{02} (2026), 021
doi:10.1088/1475-7516/2026/02/021
[\hhref{2507.21215}].



%\cite{Gibbons:1977mu}
\bibitem{Gibbons:1977mu}
G.~W.~Gibbons and S.~W.~Hawking,
``Cosmological Event Horizons, Thermodynamics, and Particle Creation,''
Phys. Rev. D \textbf{15}, 2738-2751 (1977)
doi:10.1103/PhysRevD.15.2738


%\cite{Caprini:2015zlo}
\bibitem{Caprini:2015zlo}
C.~Caprini, M.~Hindmarsh, S.~Huber, T.~Konstandin, J.~Kozaczuk, G.~Nardini, J.~M.~No, A.~Petiteau, P.~Schwaller, G.~Servant and D.~J.~Weir,
``Science with the space-based interferometer eLISA. II: GWs from cosmological phase transitions,''
JCAP \textbf{04} (2016), 001
doi:10.1088/1475-7516/2016/04/001
[\hhref{1512.06239}].

%\cite{Caprini:2018mtu}
\bibitem{Caprini:2018mtu}
C.~Caprini and D.~G.~Figueroa,
``Cosmological Backgrounds of Gravitational Waves,''
Class. Quant. Grav. \textbf{35} (2018) no.16, 163001
doi:10.1088/1361-6382/aac608
[\hhref{1801.04268}].

%\cite{vonHarling:2019gme}
\bibitem{vonHarling:2019gme}
B.~Von Harling, A.~Pomarol, O.~Pujol\`as and F.~Rompineve,
``Peccei-Quinn Phase Transition at LIGO,''
JHEP \textbf{04} (2020), 195
doi:10.1007/JHEP04(2020)195
[\hhref{1912.07587}].

%\cite{Coleman:1977th}
\bibitem{Coleman:1977th}
S.~R.~Coleman, V.~Glaser and A.~Martin,
``Action Minima Among Solutions to a Class of Euclidean Scalar Field Equations,''
Commun. Math. Phys. \textbf{58} (1978), 211-221 doi:10.1007/BF01609421

\bibitem{Brezin-Parisi} E. Brezin and G. Parisi, J. Stat. Phys. 19 (1978) 269 doi:10.1007/BF01011726

%\cite{Arnold:1991cv}
\bibitem{Arnold:1991cv}
P.~B.~Arnold and S.~Vokos,
``Instability of hot electroweak theory: bounds on m(H) and M(t),''
Phys. Rev. D \textbf{44} (1991), 3620-3627
doi:10.1103/PhysRevD.44.3620

\bibitem{MezoBaricz} I.~Mez\H{o} and \'A.~Baricz, ``On the generalization of the Lambert W function,'' Trans.\ Amer.\ Math.\ Soc.\ {\bf 369} (2017) 7917 [\hhref{1408.3999}].

	
\bibitem{dataset}
F.~Cutrona and A.~Salvio, ``On the Role of Gravity in Supercooled Phase Transitions; Dataset"   (2026) https://doi.org/10.5281/zenodo.23734748.

%\cite{Boyle:2007zx}
\bibitem{Boyle:2007zx}
L.~A.~Boyle and A.~Buonanno,
``Relating gravitational wave constraints from primordial nucleosynthesis, pulsar timing, laser interferometers, and the CMB: Implications for the early Universe,''
Phys. Rev. D \textbf{78} (2008), 043531
doi:10.1103/PhysRevD.78.043531
[\hhref{0708.2279}].

%\cite{Stewart:2007fu}
\bibitem{Stewart:2007fu}
A.~Stewart and R.~Brandenberger,
``Observational Constraints on Theories with a Blue Spectrum of Tensor Modes,''
JCAP \textbf{08} (2008), 012
doi:10.1088/1475-7516/2008/08/012
[\hhref{0711.4602}].

%\cite{Kohri:2018awv}
\bibitem{Kohri:2018awv}
K.~Kohri and T.~Terada,
``Semianalytic calculation of gravitational wave spectrum nonlinearly induced from primordial curvature perturbations,''
Phys. Rev. D \textbf{97} (2018) no.12, 123532
doi:10.1103/PhysRevD.97.123532
[\hhref{1804.08577}].

%\cite{LIGOScientific:2026mjf}
\bibitem{LIGOScientific:2026mjf}
A.~G.~Abac \textit{et al.} [LIGO Scientific, VIRGO and KAGRA],
``Updated Upper Limits on the Isotropic Gravitational-Wave Background from LIGO, Virgo, and KAGRA Data through April 2025,''
[\hhref{2608.23477}].

\bibitem{LIGOO5}
LIGO Scientific Collaboration, ``Updated Advanced LIGO sensitivity design curves for O5,"
LIGO Document T2500310 (2025), https://dcc.ligo.org/LIGO-T2500310/public.

%\cite{KAGRA:2021kbb}
\bibitem{KAGRA:2021kbb}
R.~Abbott \textit{et al.} [KAGRA, Virgo and LIGO Scientific],
``Upper limits on the isotropic gravitational-wave background from Advanced LIGO and Advanced Virgo\textquoteright{}s third observing run,''
Phys. Rev. D \textbf{104} (2021) no.2, 022004
doi:10.1103/PhysRevD.104.022004
[\hhref{2101.12130}].

%\cite{LIGOScientific:2023vdi}
\bibitem{LIGOScientific:2023vdi}
R.~Abbott \textit{et al.} [LIGO Scientific, VIRGO and KAGRA],
``Open data from the third observing run of LIGO, Virgo, KAGRA and GEO,''
[\hhref{2302.03676}].


%\cite{Babak:2021mhe}
\bibitem{Babak:2021mhe}
S.~Babak, A.~Petiteau and M.~Hewitson,
``LISA Sensitivity and SNR Calculations,''
[\hhref{2108.01167}].

%\cite{Evans:2021gyd}
\bibitem{Evans:2021gyd}
M.~Evans, R.~X.~Adhikari, C.~Afle, S.~W.~Ballmer, S.~Biscoveanu, S.~Borhanian, D.~A.~Brown, Y.~Chen, R.~Eisenstein and A.~Gruson, \textit{et al.}
``A Horizon Study for Cosmic Explorer: Science, Observatories, and Community,''
[\hhref{2109.09882}].

%\cite{Thrane:2013oya}
\bibitem{Thrane:2013oya}
E.~Thrane and J.~D.~Romano,
``Sensitivity curves for searches for gravitational-wave backgrounds,''
Phys. Rev. D \textbf{88} (2013) no.12, 124032
[\hhref{1310.5300}].

%\cite{Dev:2019njv}
\bibitem{Dev:2019njv}
P.~S.~B.~Dev, F.~Ferrer, Y.~Zhang and Y.~Zhang,
``Gravitational Waves from First-Order Phase Transition in a Simple Axion-Like Particle Model,''
JCAP \textbf{11} (2019), 006
[\hhref{1905.00891}].

%\cite{Buttazzo:2013uya}
\bibitem{Buttazzo:2013uya}
D.~Buttazzo, G.~Degrassi, P.~P.~Giardino, G.~F.~Giudice, F.~Sala, A.~Salvio and A.~Strumia,
``Investigating the near-criticality of the Higgs boson,''
JHEP \textbf{12}, 089 (2013)
doi:10.1007/JHEP12(2013)089
[\hhref{1307.3536}].


  
  }
  

\end{thebibliography}
\end{document}